\documentclass[3p]{elsarticle}

\usepackage[utf8]{inputenc}

\usepackage{amssymb}
\usepackage{epsfig}
\usepackage{comment}
\usepackage{amsmath}
\usepackage[section]{placeins}
\usepackage{mathrsfs}
\usepackage{graphicx}
\usepackage[notrig]{physics}
\usepackage{units}
\usepackage{subfigure}
\usepackage{caption}
\usepackage{color}
\usepackage{svg}
\usepackage[hidelinks]{hyperref}
 
\journal{Computer Methods in Applied Mechanics and Engineering}

\newtheorem{thm}{Theorem}[section]
\newtheorem{example}[thm]{Example}

\usepackage{stackengine}
\setstackgap{S}{1.75pt}

\renewcommand{\appendixname}{Appendix\hspace{0.1em}}

\newcommand\tB[1]{{\color{black}#1}}
\definecolor{myblue}{RGB}{0, 0, 225}

\biboptions{authoryear}

\begin{document}

\title{Homogenized Models of Mechanical Metamaterials$^{\dagger}$}

\author[1]{J.~Ulloa\corref{cor1}}
\ead{julloa@caltech.edu}

\author[2]{M.~P.~Ariza}
\ead{mpariza@us.es}

\author[1]{J.~E.~Andrade}
\ead{jandrade@caltech.edu}

\author[1]{M.~Ortiz}

\ead{ortiz@caltech.edu}

\cortext[cor1]{Corresponding author}

\affiliation[1]{organization={Division of Engineering and Applied Science, California Institute of Technology},
                 addressline={1200 E.~California Blvd.},
                 city={Pasadena},
                 postcode={CA 91125},
                 country={USA}}
\affiliation[2]{organization={Escuela Técnica Superior de Ingeniería, Universidad de Sevilla},
                 addressline={Camino de los descubrimientos, s.n.},
                 postcode={41092},
                 city={Sevilla},
                 country={Spain}}

\date{\today}

\begin{abstract}
Direct numerical simulations of mechanical metamaterials are prohibitively expensive due to the separation of scales between the lattice and the macrostructural size. Hence, multiscale continuum analysis plays a pivotal role in the computational modeling of metastructures at macroscopic scales. In the present work, we assess the continuum limit of mechanical metamaterials via homogenized models derived rigorously from variational methods. It is shown through multiple examples that micropolar-type effective energies, derived naturally from analysis, properly capture the kinematics of discrete lattices in two and three dimensions. Moreover, the convergence of the discrete energy to the continuum limit is shown numerically. We provide open-source computational implementations for all examples, including both discrete and homogenized models.
\end{abstract}

\maketitle

\section{Introduction}
{\sl Architected materials}, or {\sl metamaterials}, exhibit remarkable mechanical properties while providing opportunities for efficient and lightweight material use~\citep{Montemayor2015,Greer:2019,LU2022,Jin2024}. Consequently, mechanical metamaterials have received significant attention over the past couple of decades, perhaps starting with the predecessor concept of {\sl cellular solids}~\citep{deshpande2001a,Ashby2006,Fleck2010}. In particular, the reticular nature of metamaterials provides great flexibility for designing architected microstructures with tailored properties, ranging from unusual density, stiffness, Poisson's ratio or strength to shape-morphing capabilities~\citep{ashby2011a,bertoldi2017}. Numerous studies have thus focused on characterizing macroscopic structural properties arising from various microstructural configurations. Examples of microstructures include the honeycomb lattice~\citep{CHEN1998,bauer2014a,kuszczak2023} and the octet truss~\citep{deshpande2001b,meza2015a,gu2015a}, among others~\citep{Vigliotti2012,rys2014a,ros2015a,bauer2016a}. 

Given the separation of scales between microstructural features and engineering components, be it strict or non-strict, full-resolution models incorporating the discrete microstructural lattice are prohibitively expensive from a computational standpoint. This situation is an exemplary case of multiscale problems, where performing continuum model simulations requires knowledge of the effective mechanical properties of the microstructure. The passing of information from the lattice to the continuum scale can be achieved in many ways in the context of multiscale analysis (see, e.g., \cite{kochmann:2019} for an overview). Examples include computational techniques such as coarse-graining \citep{PHLIPOT2019} or FE$^2$ methods~\citep{danesh2023}. More recently, data-based approaches have also emerged, either through machine learning techniques \citep{zhang2024} or data-driven mechanics \citep{Weinberg:2023}. Of course, in these computational approaches, an underlying functional form of the effective model and the associated material parameters are not identified in closed form. 

Alternatively, effective macroscopic properties may be identified analytically, particularly for elastic materials, in terms of the lattice geometry and base material properties~\citep{CHEN1998,deshpande2001a}. In this context, \cite{Ariza:2024} recently developed a homogenization framework based on the discrete-to-continuum method within the calculus of variations~\citep{Cicalese:2009,Braides:2004}. Remarkably, this approach is provably convergent in energy and minimizing solutions. Furthermore, it allows us to identify unambiguously the functional form and all effective properties in an ansatz-free manner, i.e., without resorting to ad hoc considerations. 

Mechanical metamaterials are subject to joint rotations and bar bending. Hence, the discrete-to-continuum analysis of~\cite{Ariza:2024} is non-standard. In fact, assuming a strict separation of scales, the zeroth-order continuum limit of linear metamaterials under axial and bending deformations was found to be micropolar, in the sense of~\cite{Eringen:1964}, albeit independent of the curvature strain, i.e., without a notion of length scale. Moreover, closed-form expressions for the elastic and micropolar moduli were obtained for different microstructures as a function of the geometry and both the axial and bending stiffness of the bars. In a separate work~\citep{Ulloa:2024}, we have further elaborated a second-order continuum limit, which retains unit-cell information, i.e., a length scale parameter, and allows us to capture size effects observed experimentally~\citep{Shaikeea2022} in particular structural~set-ups.

\tB{In the present work, we provide a first and thorough computational assessment of the convergence of discrete numerical solutions to the variational continuum limit~\citep{Ariza:2024}. Accordingly, our continuum models are established from variational principles without postulating a kinematic ansatz, in sharp contrast with previous computational studies on generalized continua for mechanical metamaterials~\citep{dos2012construction, rokovs2019micromorphic, neff2020identification, misra2020chiral, biswas2020micromorphic, alavi2021construction}.} We consider a variety of two- and three-dimensional test cases to assess the energy convergence numerically. The calculations verify the analytical results and exhibit clear convergence rates, depending on the specific structural setting. The finite element implementation of the discrete and homogenized models is provided as open-source code in {\tt MATLAB} and~{\tt Python}.

The paper is structured as follows. Section~\ref{sec:meta} reviews the discrete formulation of mechanical metamaterials, setting the stage for the homogenization analysis described in Sec.~\ref{sec:homog}. Then, Sec.~\ref{sec:comp} presents a series of numerical examples considering honeycomb lattices in two dimensions and octet trusses in three dimensions. Conclusions and perspectives are finally given in Sec.~\ref{sec:conc}.

\section{Metamaterials}\label{sec:meta}

\tB{A broad class of mechanical metamaterials can be described as locally periodic frame structures, consisting of {\sl bars} connected at a set of {\sl joints} (cf., e.g., \cite{Fleck2010, Montemayor2015})}. In this section, we summarize a discrete mechanics formulation of metamaterials \citep{Ariza:2024} that provides a suitable basis for analysis and computation. 

\subsection{Geometry of metamaterials}
\label{seckinematics}

\begin{figure}[ht!]
\begin{center}
    \subfigure[]{\includegraphics[width=0.5\textwidth]{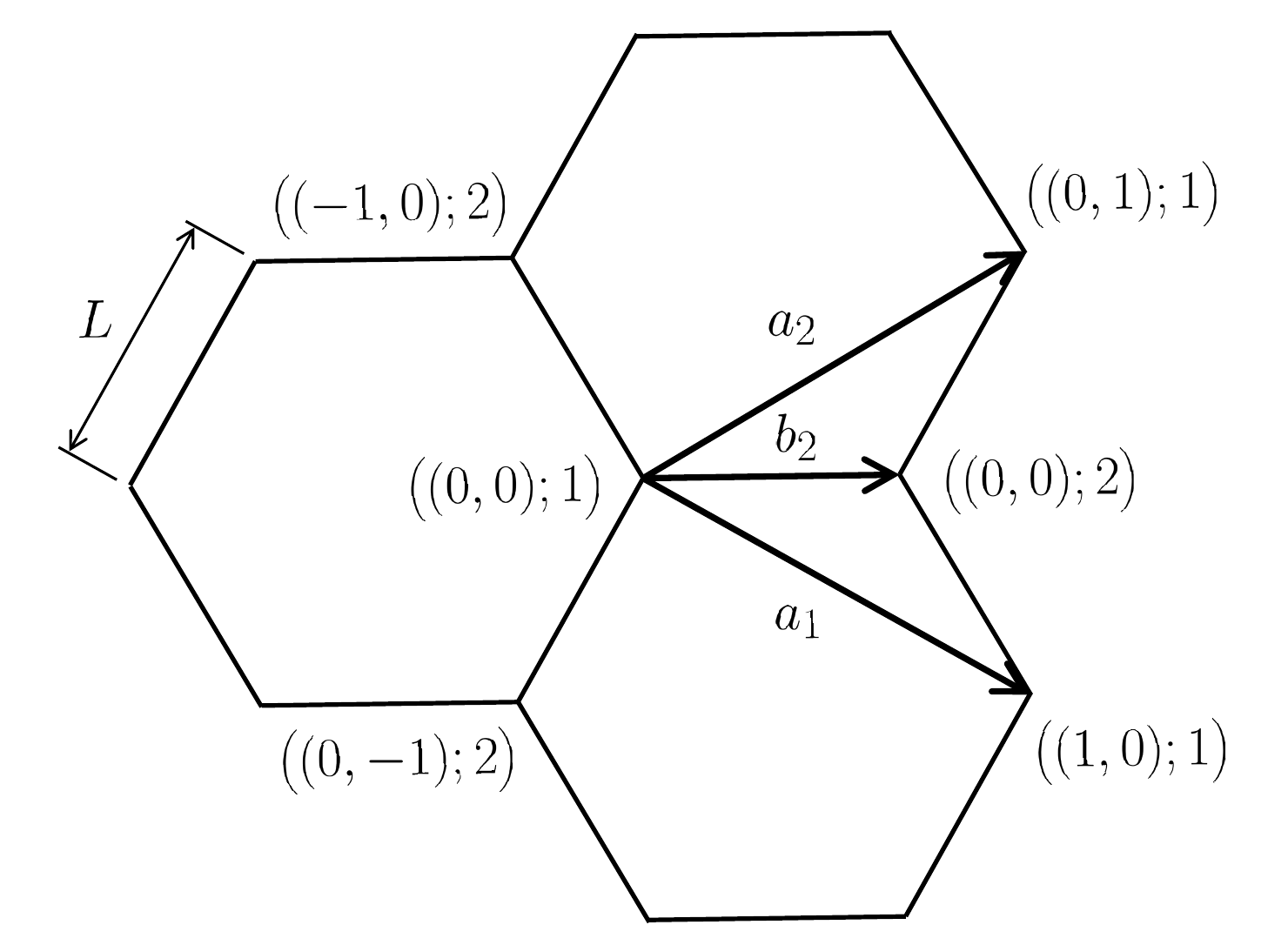}}
    \subfigure[]{\includegraphics[width=0.39\textwidth]{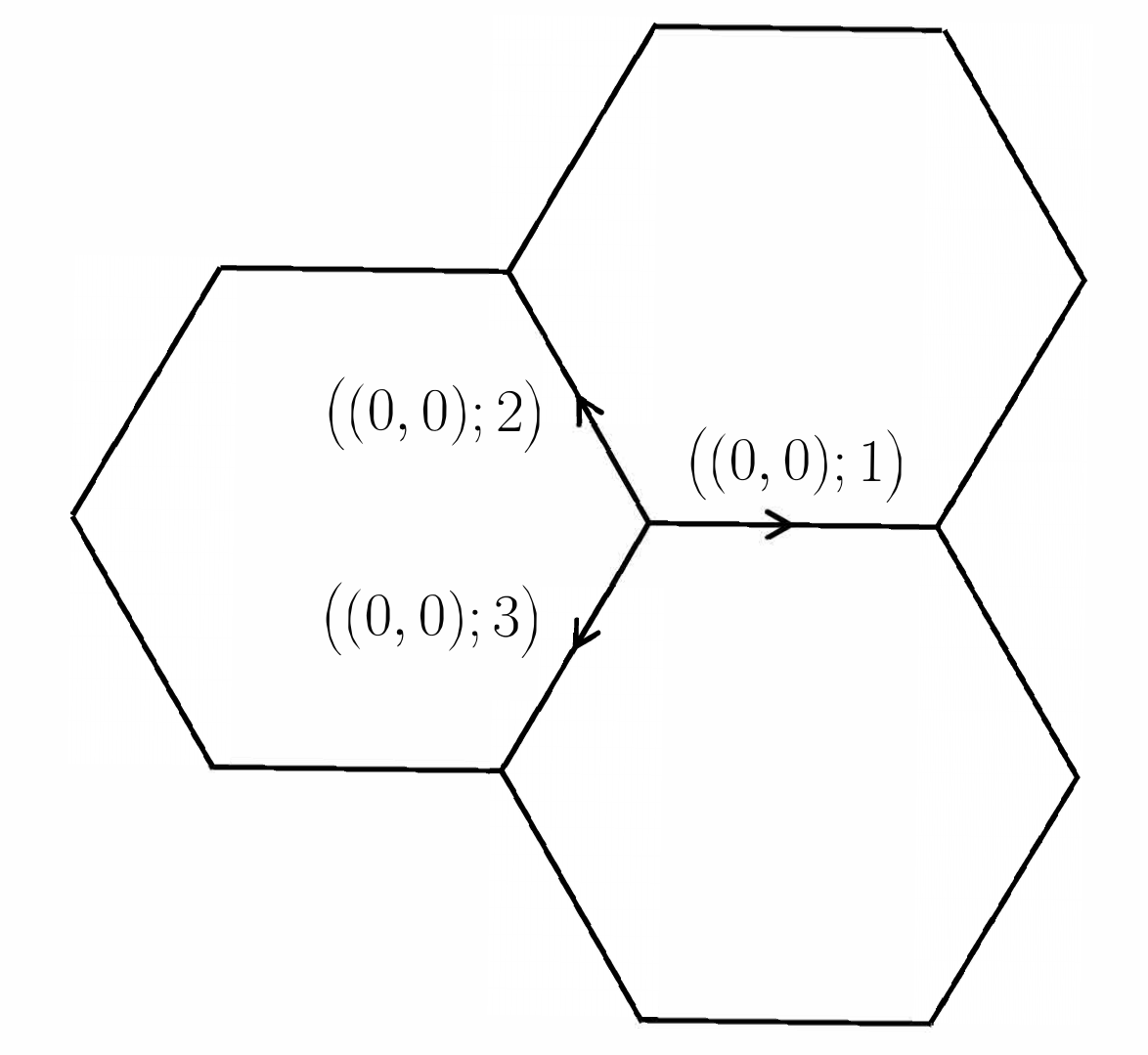}}
    \caption{\tB{Honeycomb metamaterial. (a) Joint numbering scheme using two simple Bravais sublattices ($N=2$), where $((l^1,l^2);\alpha)$ are the joint indices, with $\alpha\in\{1,2\}$; $a_1 = L\,(3/2,-\sqrt{3}/2)$ and $a_2 = L\,(3/2,\sqrt{3}/2)$ are the two-dimensional basis vectors; and $b_1=(0,0)$ and $b_2=L\,(1,0)$ are the translation vectors. (b) Member numbering scheme using three simple Bravais sublattices ($M=3$), where $((m^1,m^2);\beta)$ are the bar indices, with $\beta\in\{1,2,3\}$.}} \label{FIDpWS}
\end{center}
\end{figure}

\begin{figure*}[h!]
\begin{center}
    \subfigure[]{\includegraphics[width=0.8\textwidth]{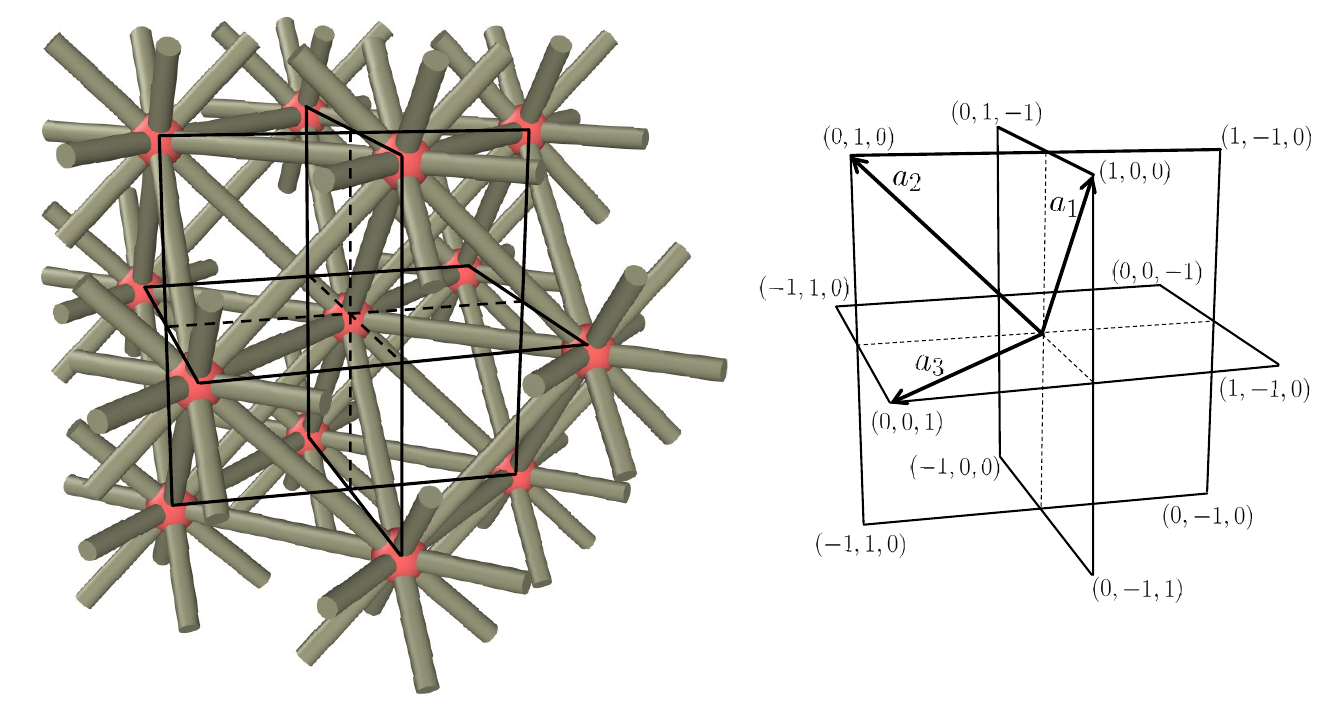}}
    \subfigure[]{\includegraphics[width=0.7\textwidth]{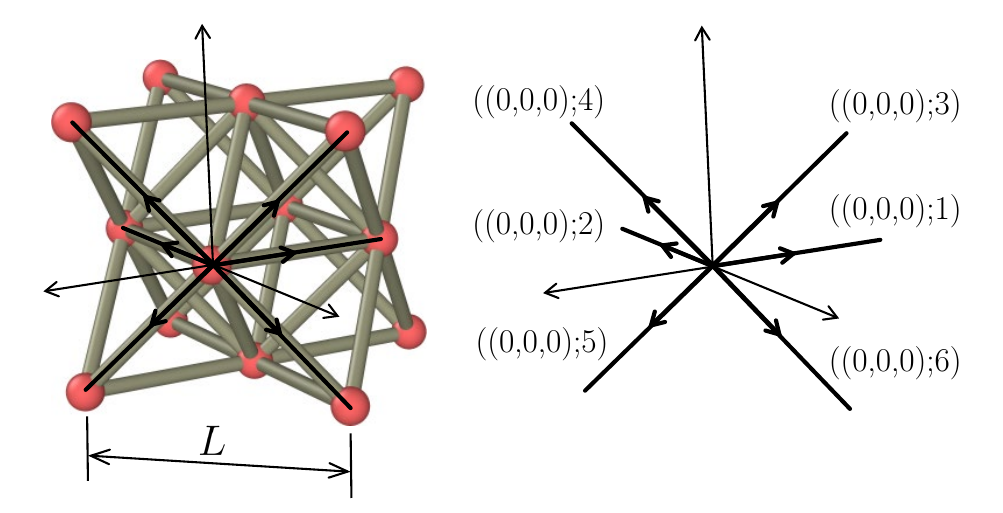}}
    \caption{\tB{Octet-truss metamaterial. (a) Joint numbering scheme using a simple Bravais lattice ($N=1$), where $(l^1,l^2,l^3)$ are the joint indices, with the single joint class index $\alpha=1$ omitted; $a_1 = (L/\sqrt{2})\,(0,1,1)$, $a_2 =(L/\sqrt{2})\,(1,0,1)$, and $a_3 = (L/\sqrt{2})\,(1,1,0)$ are the three-dimensional basis vectors; and $b_1=(0,0,0)$. (b) Bar numbering scheme ($M=6$), where $((m^1,m^2,m^3);\beta)$ are the bar indices, with $\beta\in\{1,2,3,4,5,6\}$.}} \label{S2rFkk}
\end{center}
\end{figure*}

Metamaterials are comprised of $M$ types, or {\sl classes}, of oriented bars, all the bars in a class being identical, including orientation, modulo translations. For instance, the bar classes of a honeycomb metamaterial are $\{\diagdown,\, \text{---},\, \diagup \}$, or $M=3$, Fig.~\ref{FIDpWS}. \tB{On the other hand, the octet-truss lattice consists of six bar classes, $M=6$, Fig.~\ref{S2rFkk}}. The {\sl local environment} of a joint is the set of bars incident on the joint. Metamaterials contain $N$ different types, or classes, of local environments, all local environments in a class being identical modulo translations. Correspondingly, the joints of the metamaterial are classified into $N$ types, according to their local environment. For instance, the joint classes of a honeycomb metamaterial are $\{\rotatebox[origin=c]{90}{${\textsf{Y}}$},\, \rotatebox[origin=c]{-90}{${\textsf{Y}}$}\}$, or $N=2$, Fig.~\ref{FIDpWS}. \tB{On the other hand, an octet-truss presents a single joint type, $N=1$, Fig.~\ref{S2rFkk}}.

The joints and bars of a metamaterial are embedded periodically in space so that joints and bars of the same class span shifted {\sl Bravais lattices}. Thus, the positions of the joints in each class span a point set of the form
\begin{equation} \label{M6JjBD}
  x(l,\alpha) = b_\alpha + \sum_{i=1}^n l^i a_i ,
  \quad l \in \mathbb{Z}^n , \quad \alpha\in\{1,\dots,N\} ,
\end{equation}
where $n$ is the dimension of space, e.g., $n=2$ for two-dimensional lattices and $n=3$ for three-dimensional lattices; $(a_i)_{i=1}^n$ is a {\sl basis} of $\mathbb{R}^n$; $l = (l^i)_{i=1}^n$ is a corresponding array of integer {\sl lattice coordinates}; and $(b_\alpha)_{\alpha=1}^N$ are translation vectors, or {\sl shifts}. We additionally denote by $V$ the volume of the unit cell of the spanning Bravais lattice of the metamaterial. The joints of a metamaterial can therefore be {\sl indexed} by a pair $(l,\alpha)$, where $\alpha \in \{1,\dots,N\}$ designates the joint class and $l \in \mathbb{Z}^n$ its lattice coordinate array. 

A full description of a metamaterial additionally requires a scheme for indexing its bars by type and position, and for tabulating their connections to joints. By periodicity and translation invariance, the bars of a given bar class span the same Bravais lattice as the joint classes, modulo translations. Thus, bars can be indexed by pairs $(m,\beta)$, where $\beta \in\{ 1,\dots,M \}$ designates the bar class and $m \in \mathbb{Z}^n$ is a lattice coordinate array. In addition, an {\sl orientation} of the bar classes identifies beginning and end joints in every bar, which we designate by the symbols $\pm$, respectively.

The adjacency relations between bars and joints, or {\sl connectivity}, can be defined by designating, for every bar $(m,\beta)$, the corresponding beginning and end joints, $(m^-,\beta^-)$ and $(m^+,\beta^+)$, respectively. The corresponding joint coordinates follow from \eqref{M6JjBD} as
\begin{equation} \label{46x3dB}
    x^\pm(m,\beta) 
    =
    x(m^\pm,\beta^\pm)
    =
    b_{\beta^\pm} + \sum_{i=1}^n (m^\pm)^i a_i ,
\end{equation}
and the {\sl spanning vector} of the joints follows as
\begin{equation}
    dx_\beta
    =
    x^+(m,\beta) - x^-(m,\beta) ,
\end{equation}
which is independent of $m$ by translation invariance. We also denote by $[x^-(m,\beta),\, x^+(m,\beta)]$ the spanning segments of the bars, regarded as sets.

\subsection{Elastic energy} \label{7PHmId}

The extent of a metastructure, i.e., its set of bars, may be characterized by means of an index set
\begin{equation} \label{0Jdwod}
\begin{split}
    &
    J
    =
    \{ (m,\beta) \in \mathbb{Z}^n\times \{1,\dots, M\} \, : \, 
    \text{bar} \; (m,\beta) \; \text{in metastructure} \} .
\end{split}
\end{equation}
For an infinite metamaterial, $J = \mathbb{Z}^n\times \{1,\dots, M\}$. Correspondingly, the index set
\begin{equation} 
\begin{split}
    &
    I
    =
    \{ (l,\alpha) \in \mathbb{Z}^n\times \{1,\dots, N\} \, : \, 
    \text{joint} \; (l,\alpha) \; \text{in metastructure} \} 
\end{split}
\end{equation}
designates the set of joints in the metastructure. For an infinite metamaterial, $I = \mathbb{Z}^n\times \{1,\dots, N\}$. 

Within the framework of linearized kinematics, the joints of a metastructure are endowed with deflection and rotation-angle degrees of freedom, $v(l,\alpha) \in \mathbb{R}^n$ and $\theta(l,\alpha) \in \mathbb{R}^{n(n-1)/2}$, respectively. We further denote by 
\begin{equation} \label{FFPJsh}
  u(l,\alpha) = (v(l,\alpha); \theta(l,\alpha)) 
\end{equation}
the set of degrees of freedom of joint $(l,\alpha)$.  

At the local level, we denote by
\begin{equation} \label{0MjvOi}
  {U}(m,\beta) = (u^-(m,\beta); u^+(m,\beta)) 
\end{equation}
the set of local degrees of freedom of bar $(m,\beta)$, where $u^\pm(m,\beta)$ are the degree-of-freedom arrays of the joints of the bar. An application of elementary beam theory \citep{Timoshenko:1965} then gives the total energy of a metastructure as
\begin{equation} \label{1ElWKO}
    E(u)
    =
    \sum_{(m,\beta)\in J}
    \frac{1}{2} \,
    \mathbb{S}_\beta \, {U}(m,\beta) \cdot {U}(m,\beta),
\end{equation}
where $\mathbb{S}_\beta$ is the elastic stiffness of a bar of class $\beta$. Explicit expressions for $E(u)$ in terms of bar geometry, connectivity, and elastic properties are given in \cite{Ariza:2024}, \tB{assuming Euler-Bernoulli beams}. 

Suppose that the metastructure is acted upon by forces 
\begin{equation}
    f(l,\alpha) = \big( q(l,\alpha) ;\, m(l,\alpha) \big), 
\end{equation}
where $q(l,\alpha)$ and $m(l,\alpha)$ are the force and moment, or torque, applied to joint $(l,\alpha) \in I$, respectively. The corresponding work function is, then,
\begin{equation} 
    \langle f,\, u \rangle
    =
    \sum_{(l,\alpha)\in I} 
        f(l,\alpha) \cdot u(l,\alpha) ,
\end{equation} 
and the potential energy of the metastructure follows as
\begin{equation} \label{YGmtm4}
    F(u) = E(u) - \langle f,\, u \rangle .
\end{equation}
By the principle of minimum potential energy, the stable equilibrium configurations of the metastructure, if any, follow by minimization of $F(u)$ over all admissible displacements.

\section{Homogenization and continuum limit}\label{sec:homog}

We consider a sequence of increasingly finer metamaterials, scaled by a small parameter $\epsilon > 0$, maximally contained within a fixed domain $\Omega$, and deforming under the action of macroscopic loads $f_0(x)$, independent of $\epsilon$. We seek to ascertain the asymptotic behavior of a sequence of suitably scaled potential energies
\begin{equation} \label{1C3ijx}
    F_\epsilon(u)
    =
    E_\epsilon(u)
   -
    \langle f_0,\, u \rangle_\epsilon ,
\end{equation}
to be defined, in the limit of $\epsilon \to 0$, or {\sl continuum limit}. 

\subsection{Scaling}
In order to define the sequence of scaled energies, we designate a reference metamaterial of fixed lattice size and denote by $E(u_\epsilon; \Omega/\epsilon)$ the energy of the corresponding metastructure maximally contained in $\Omega/\epsilon$. The energies of the scaled metastructures maximally contained in $\Omega$ are, then,
\begin{equation} \label{R79mPV}
    E_\epsilon(u;\, \Omega)
    =
    \epsilon^{n} 
    \, E(u_\epsilon;\, \Omega/\epsilon) ,
\end{equation}
with
\begin{equation} \label{IAuati}
    x_\epsilon =  \epsilon^{-1} \, x,
    \quad
    v_\epsilon(x_\epsilon) 
    =
    \epsilon^{-1} \, v(x) ,
    \quad
    \theta_\epsilon(x_\epsilon)
    = \theta(x) .
\end{equation}
This scaling is chosen so that both strains and rotations remain $O(1)$ as $\epsilon \downarrow 0$. In addition, the factor $\epsilon^{n}$ accounts for the expected elasticity scaling of the limiting energy and is included to ensure a proper limit \citep{Espanol:2013, Ariza:2024}. 

Suppose, in addition, that the sequence of scaled metastructures is acted upon by macroscopic distributed loads per unit volume, independent of $\epsilon$, represented by continuous functions
\begin{equation}
    f_0(x) = \big( q_0(x) ;\, m_0(x) \big) ,
\end{equation}
where $q_0(x)$ and $m_0(x)$ are distributed forces and moments, or torques, per unit volume, respectively. The corresponding joint loads on the $\epsilon$-metamaterial are
\begin{equation} \label{7ZUaPg}
    f(l,\alpha)
    =
    \frac{V_\epsilon}{N}
    f_0(x(l,\alpha)) , 
\end{equation}
for all joints contained in $\Omega$, where $V_\epsilon= \epsilon^n V$ is the volume of the unit cell in the scaled Bravais lattice, $V$ is the volume of the unit cell in the reference Bravais lattice, and $V_\epsilon/N$ is the corresponding volume per joint class. The work of the applied actions is, then,
\begin{equation} \label{XVOWv4}
    \langle f_0,\, u \rangle_\epsilon
    =
    \frac{V_\epsilon}{N} \, 
    \sum_{(l,\alpha) \in I_\epsilon}
        f_0(x(l,\alpha)) \cdot u(l,\alpha) ,
\end{equation} 
where $I_\epsilon$ is the joint index set of the $\epsilon$-metamaterial maximally contained in $\Omega$.

\subsection{Fourier representation}

In the special case of infinite metamaterials, periodicity can be exploited to represent the discrete displacements in terms of their discrete Fourier transform,
\begin{equation} \label{m0AagV}
    u(l,\alpha)
    =
    \frac{1}{(2\pi)^n}
    \int_B
        \hat{u}(k,\alpha) \, {\rm e}^{i k \cdot x(l,\alpha)}
    \, dk ,
\end{equation}
where $B$ is the Brillouin zone of the metastructure, $x(l,\alpha)$ are the joint coordinates \eqref{M6JjBD}, and $\hat{u}(k,\alpha)$ is the discrete Fourier transform of the discrete displacement field $u(l,\alpha)$, cf.~\ref{TwWqmj}.

The energy then takes the compact form 
\begin{equation} \label{oEf3Zx}
    E(u)
    =
    \frac{1}{(2\pi)^n}
    \int_B
        \frac{1}{2}
        \mathbb{D}(k) 
        \, \hat{u}(k) \cdot \hat{u}^*(k) 
    \, dk   
\end{equation}
in terms of the {\sl dynamical matrix} $\mathbb{D}(k)$ of the metamaterial. Here, we have collected all joint-type displacements in the Fourier representation \eqref{m0AagV} into a single array, not renamed,
\begin{equation} \label{qbNEqL1}
    \hat{u}(k) = \big( \hat{u}(k,1), \dots, \hat{u}(k,N) \big) .
\end{equation}
We note that $\mathbb{D}(k)$ is Hermitian, positive definite, depends smoothly on the wavevector $k$, and is supported in the Brillouin zone $B$. Explicit expressions for $\mathbb{D}(k)$ in terms of bar geometry, connectivity, and elastic properties are presented in \cite{Ariza:2024}. 

In light of the Fourier representation~\eqref{oEf3Zx}, the scaling \eqref{R79mPV}--\eqref{IAuati} defines a sequence of scaled energies of the form
\begin{equation} \label{tt54ez}
    E_\epsilon(u)
    =
    \frac{1}{(2\pi)^n}
    \int_{B/\epsilon}
        \frac{1}{2}
        \mathbb{D}_\epsilon(k) 
        \, \hat{u}(k) \cdot \hat{u}^*(k) 
    \, dk ,    
\end{equation}
where the scaled dynamical matrix $\mathbb{D}_\epsilon(k)$ is defined by the property
\begin{equation} \label{3ll62m}
\begin{split}
  &
  \mathbb{D}_\epsilon(k) \, \zeta \cdot \zeta^* 
  = 
  \epsilon^{-2} \,
  \mathbb{D}(\epsilon k) \, 
    (\xi; \, \epsilon \eta)
    \cdot
    (\xi^*; \, \epsilon \eta^*) ,
  \\ & 
  \forall \zeta = (\xi;\,\eta) \in [\mathbb{C}^n]^N \times [\mathbb{C}^{n(n-1)/2}]^N ,
  \quad
  k \in B/\epsilon .
\end{split}
\end{equation}
Likewise, the work function \eqref{XVOWv4} becomes
\begin{equation} \label{W8oi93}
    \langle f_0,\, u \rangle_\epsilon
    :=
    \frac{1}{(2\pi)^n}
    \int_{B/\epsilon}
        \hat{f}(k) \cdot \hat{u}_\epsilon^*(k)
    \, dk ,
\end{equation} 
where we write
\begin{equation} \label{qbNEqL}
    \hat{f}(k)
    =
    \Big( 
        \frac{1}{N}
        \hat{f}_0(k), 
        \stackrel{N}{\dots}, 
        \frac{1}{N}
        \hat{f}_0(k) 
    \Big) 
    := 
    L \hat{f}_0(k) .
\end{equation}
We note that the operator $L$ thus defined {\sl localizes} the continuum forces to each of the joint classes. 

Assuming equicoercivity of the sequence of scaled energies ($E_\epsilon$), the scaled dynamical matrix $\mathbb{D}_\epsilon(k)$ is non-singular, as required for structural stability. We then find the minimizing generalized displacements pointwise in $B/\epsilon$ as
\begin{equation} \label{jQ2xKs}
    \hat{u}_\epsilon(k)
    =
    \mathbb{D}_\epsilon^{-1}(k) \, \hat{f}(k) ,
    \quad
    k \in B/\epsilon .
\end{equation}
Replacing this expression in~\eqref{W8oi93},~\eqref{tt54ez}, and~\eqref{1C3ijx} yields the corresponding minimum potential energy
\begin{equation}\label{meps}
    m_\epsilon(E_\epsilon;f_0)
    := F_\epsilon(u_\epsilon)
    =
    -
    \frac{1}{(2\pi)^n}
    \int_{B/\epsilon}
        \frac{1}{2}
        \mathbb{D}_\epsilon^{-1}(k) 
        \, \hat{f}(k) \cdot \hat{f}^*(k)
    \, dk .
\end{equation}

\subsection{Homogenization limit}

We wish to determine the variational limit $E_0$ of the sequence $(E_\epsilon)$ as $\epsilon \to 0$, in the sense of $\Gamma$-convergence. This limit is characterized by the property that the minimum potential energy~\eqref{meps} converges to the minimum $m_0(E_0, f_0)$ of the limiting potential energy
\begin{equation} 
    F_0(u_0)
    =
    E_0(u_0)
    -
    \langle f_0,\, u_0 \rangle
\end{equation}
as $\epsilon \to 0$ {\sl for all applied loads} $f_0(x)$. More specifically,
\begin{equation} \label{7rqERh}
    \lim_{\epsilon\to 0} m_\epsilon(E_\epsilon, f_0) = m_0(E_0, f_0) =  -E_0^*(f_0),
\end{equation}
where $E_0^*(f_0)$ is the {\sl dual} effective energy.

This limit is greatly simplified in the case of linear metastructures, wherein the total energy is a quadratic form (see, e.g., \cite{dalmaso:1993} for background on $\Gamma$- and $G$-convergence of quadratic forms). The {\sl primal} effective energy follows from~\eqref{7rqERh} as \citep{Ariza:2024}
\begin{equation} \label{lIKdde}
    E_0(u_0)
    =
    \int_{\Omega}
        W_0\big(\varepsilon_0(x),\, 
        \theta_0(x) - \tfrac{1}{2} \operatorname{curl} v_0(x)\big)
    \, dx ,
\end{equation}
where $\varepsilon_0(x) = {\rm sym}\, Dv_0(x)$ are the local strains and the quadratic function $W_0(\varepsilon,\, \omega)$, $(\varepsilon,\, \omega) \in \mathbb{R}^{n\times n} \times \mathbb{R}^{n(n-1)/2}$, is the effective energy density of the infinite metamaterial. 

Thus, to lowest order, the effective continuum energy \eqref{lIKdde} of linear metastructures is a special case of linear {\sl micropolar elasticity}, in the sense of \cite{Eringen:1966}, in which the energy density is independent of the curvature, or bending strain, $D\theta_0(x)$. Explicit expressions for the attendant effective moduli of two-dimensional honeycomb lattices and three-dimensional octet trusses are presented in \cite{Ariza:2024}. 

\begin{example}[Honeycomb lattice]\label{ex_honeycomb}
{\rm The honeycomb lattice of size $L$ is two-dimensional, $n=2$, and contains two types of joints, $N=2$, and three types of oriented bars, $M=3$. A direct computation of the limit \eqref{lIKdde} gives, to zeroth order, the limiting continuum energy \citep{Ariza:2024}
\begin{equation}\label{eq:E0hon}
    E_0(u_0) 
    = 
    \int_\Omega
        \frac{1}{2}
        \mathbb{C} \,
        \varepsilon_0(x)
        :
        \varepsilon_0(x)
    \, dx
    +
    \int_\Omega
        \frac{1}{2}
        \frac{8\sqrt{3} EI}{L^3}
        \Big( \theta_0(x) - \frac{1}{2} \operatorname{curl} v_0(x) \Big)^2
    \, dx,
\end{equation}
in terms of elastic and micropolar components. The effective elastic moduli defining the former have cubic symmetry and evaluate, in Voigt form, to
\begin{equation}
\begin{split}
    &
    C_{11}
    =
    \frac
    {
        {EA} \, \left(\, {EA} \, L^2 + 36 \, {EI} \, \right)
    }
    {
        2 \sqrt{3} \, \left(\, {EA} \, L^3 + 12 \, {EI} \, L\right)
    } ,
    \\ &
    C_{12}
    =
    \frac
    {
        {EA} \, \left({EA} \, L^2 - 12 \, {EI} \, \right)
    }
    {
        2 \sqrt{3} \, 
        \left(\, {EA} \, L^3 + 12 \, {EI} \, L\right)
    } ,
    \\ &
    C_{33}
    =
    \frac
    {
        4 \sqrt{3} \, {EA} \, {EI}
    }
    {
        {EA} \, L^3 + 12 \, {EI} \, L
    } .
\end{split}
\end{equation}
We note that the continuum energy is isotropic in the plane, as expected from the hexagonal symmetry of the lattice. 
\hfill$\square$
}
\end{example}

\begin{example}[Octet truss]\label{ex_octet}
{\rm The octet-truss metamaterial of size $L$ is three-dimensional, $n=3$, and contains one type of joints, $N=1$, and six types of oriented bars, $M=6$. 
Assuming, for simplicity, $GI_1 := GI$, $EI_2 = EI_3 := EI$, a direct computation of the limit \eqref{lIKdde} gives, to zeroth order, the limiting continuum energy \citep{Ariza:2024}
\begin{equation} \label{GQ4mav}
    E_0(u_0) 
     = 
    \int_\Omega
        \frac{1}{2}
        \mathbb{C} \,
        \varepsilon_0(x)
        :
        \varepsilon_0(x)
    \, dx + 
    \int_\Omega
        \frac{1}{2}
        \frac{48 \sqrt{2} {EI}}{L^4}
        \Big| \theta_0(x) - \frac{1}{2} \operatorname{curl} v_0(x) \Big|^2
    \, dx ,
\end{equation}
comprising an elastic and a micropolar component. The effective elastic moduli defining the former have cubic symmetry and evaluate, in Voigt form, to
\begin{equation}
\begin{split}
    &
    C_{11}
    =
    \frac{4 {EA} L^2+24 {EI}}{\sqrt{2} L^4} ,
    \\ &
    C_{12}
    =
    \frac{\sqrt{2} \left({EA} L^2-6 {EI}\right)}{L^4} ,
    \\ &
    C_{44}
    =
    \frac{2 {EA} L^2+12 {EI}}{\sqrt{2} L^4} .
\end{split}
\end{equation}
The micropolar energy density, second term in \eqref{GQ4mav}, is isotropic, as expected from the cubic symmetry of the lattice. It is also interesting to note that the effective moduli are independent of the torsional stiffness $GI_1$ of the bars. Remarkably, both axial deformations and bending contribute to the effective moduli. However, the bending terms are of order $r_g^2/L^2$, or slenderness squared, relative to the axial terms, where $r_g=\sqrt{I/A}$ is the radius of gyration of the cross-section. Therefore, the axial contribution to the effective elastic moduli is dominant for typical beam slendernesses.
\hfill$\square$
}
\end{example}

\subsection{The zeroth-order continuum boundary value problem}

\tB{With the quadratic energy density $W_0(Dv_0,\theta_0)$ in hand, we formulate the zeroth-order continuum boundary value problem as follows:  find $u_0=(v_0,\theta_0)\in U_{\bar{u}_0}$ such that
\begin{equation}
F_0(v_0,\theta_0) \leq F_0(v,\theta)  \quad \forall \ (v,\theta)\in U_{0},
\end{equation}
where
$$
U_{w} := \big\{u\in  H^1\big(\Omega;\mathbb{R}^{n}\times\mathbb{R}^{n(n-1)/2}\big) \ : \ u = w \ \text{ on } \ \Gamma_\mathrm{D}   \big\}.
$$
The first variation of $F_0$ at $u_0\in U_{\bar{u}_0}$ in the direction $u\in U_{0}$ then reads
\begin{equation}
\begin{aligned}
\langle D F_0(u_0),u \rangle = \int_\Omega \bigg\{\frac{\partial{W_0}(Dv_0,\theta_0)}{\partial Dv_0}:Dv + \frac{\partial{W_0}(Dv_0,\theta_0)}{\partial\theta_0}\cdot\theta \bigg\} \,dx  - \langle f_0,u \rangle = 0 \\ \forall \ u=(v,\theta)\in U_{0}.
\label{eq:PDE0}
\end{aligned}
\end{equation}

Let us derive for the sake of completeness the strong form of the boundary value problem. Recall that the curl operator yields in two dimensions and three dimensions the axial angle and axial vector
\begin{equation*}
\mathrm{curl}\,v = e_{\alpha\beta}\,(\mathrm{skw}\,D v)_{\beta\alpha} \quad \text{and} \quad (\mathrm{curl}\,v)_i = e_{ijk}\,(\mathrm{skw}\,D v)_{kj} 
\end{equation*}
respectively, with $e_{\alpha\beta}$ and $e_{ijk}$ the corresponding permutation tensors. The stress in the effective boundary value problem~\eqref{eq:PDE0} then admits the decomposition
\begin{equation}
\frac{\partial{W_0}(Dv_0,\theta_0)}{\partial Dv_0} = \sigma(\mathrm{sym}\,D v_0) + \tau(\mathrm{skw}\,D v_0,\theta_0),
\end{equation}
where
\begin{equation}
\sigma(\mathrm{sym}\,D v_0) = \frac{\partial{W_0}(Dv_0,\theta_0)}{\partial ( \mathrm{sym}\, Dv_0)} = \mathbb{C}\,\mathrm{sym}\,Dv_0
\end{equation}
is the standard symmetric Cauchy-type stress and
\begin{equation}
\tau(\mathrm{skw}\,D v_0,\theta_0) = \frac{\partial{W_0}(Dv_0,\theta_0)}{\partial ( \mathrm{skw}\, Dv_0)} = \frac{1}{2}\varkappa \, \big(e\cdot\theta_0-\mathrm{skw}\,Dv_0^\mathrm{T}\big)
\end{equation}
is a skew-symmetric relative stress in the micropolar sense. Here, $\varkappa$ is the coupling modulus, i.e., ${8\sqrt{3} EI}/({2L^3})$ in~\eqref{eq:E0hon} for the honeycomb lattice and ${48 \sqrt{2} {EI}}/(2{L^4})$ in~\eqref{GQ4mav} for the octet truss. On the other hand,
\begin{equation}
\frac{\partial{W_0}(Dv_0,\theta_0)}{\partial \theta_0} = \varkappa\,\big(\theta_0 - \tfrac{1}{2}\,\mathrm{curl}\,v_0 \big) = e:\tau(\mathrm{skw}\,D v_0,\theta_0)
\end{equation}
is a relative moment vector. The effective problem~\eqref{eq:PDE0} then takes the strong form
\begin{subequations}
\begin{align}
&\mathrm{div}\,\big[\sigma(\mathrm{sym}\,D v_0)  + \tau(\mathrm{skw}\,D v_0,\theta_0)\big] + q_0 = 0 &&\text{in} \quad \Omega,\\
&e:\tau(\mathrm{skw}\,D v_0,\theta_0) - m_0 = 0 &&\text{in} \quad \Omega, \\
&\big[\sigma(\mathrm{sym}\,D v_0)  + \tau(\mathrm{skw}\,D v_0,\theta_0)\big]\nu = \bar{t}_0 &&\text{on} \quad \Gamma_\mathrm{N},\\
&u_0 = \bar{u}_0 &&\text{on} \quad \Gamma_\mathrm{D},
\end{align}
\end{subequations}
where $q_0$, $m_0$, $\bar{t}_0$, and $\bar{u}_0$  are continuum body forces, body moments, imposed tractions on the Neumann boundary $\Gamma_\mathrm{N}$, and imposed generalized displacements on the Dirichlet boundary $\Gamma_\mathrm{D}$, respectively; and $\nu$ is the outer unit normal at the boundary.

}

The bilinear form~\eqref{eq:PDE0} is implemented directly for the numerical simulations that follow using standard finite element discretization. In particular, we consider quadratic interpolation for the displacement field $v_0:\Omega\to\mathbb{R}^n$ and linear interpolation for the rotation field $\theta_0:\Omega\to\mathbb{R}^{n(n-1)/2}$.

\section{Computational implementation and analysis of convergence}\label{sec:comp}

This section presents numerical results that elucidate the convergence properties of mechanical metamaterials. We seek to compare the mechanical response of discrete metastructures with the corresponding homogenized models and study the numerical convergence of the discrete energy towards the continuum limit. To this end, we consider progressively larger metastructures $\Omega/\epsilon$ with a fixed microstructure size $L$. 

We conduct direct numerical simulations for the discrete metastructures using standard Euler-Bernoulli beams, implemented in the open-source {\tt Stabil} library~\citep{franccois2021stabil}, available for {\tt MATLAB} or {\tt GNU Octave}. On the other hand, we conduct continuum finite element analysis for the homogenized models using the open-source {\tt FEniCS} library in {\tt {Python}} (cf.~\cite{suh2019open} for previous implementations of micropolar continua). The scripts for all the examples are provided as supplementary material.

\subsection{Two-dimensional honeycomb metastructures}
\label{sec:comp2D}

Let us first assess the limiting continuum model of two-dimensional honeycomb metamaterials, as derived in Example~\ref{ex_honeycomb}. To this end, we consider different boundary value problems with geometrical defects.

\begin{figure}[t!]
    \centering 
    \vspace{1em}
    \includegraphics[width=1.0\textwidth]{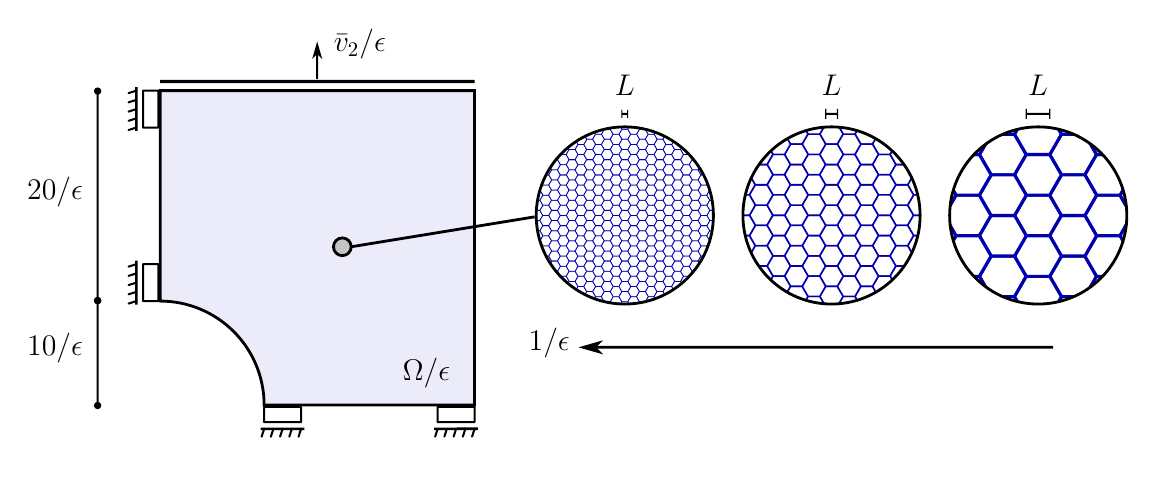}
    \caption{Boundary value problem for the perforated plate with a honeycomb microstructure. The domain is progressively scaled by $1/\epsilon$ with a fixed microstructure size $L$. Dimensions in mm.}
    \label{fig:hip_bvp}
\end{figure}

\begin{figure}[h!]
    \centering 
    \vspace{1em}
    \includegraphics[width=0.9\textwidth]{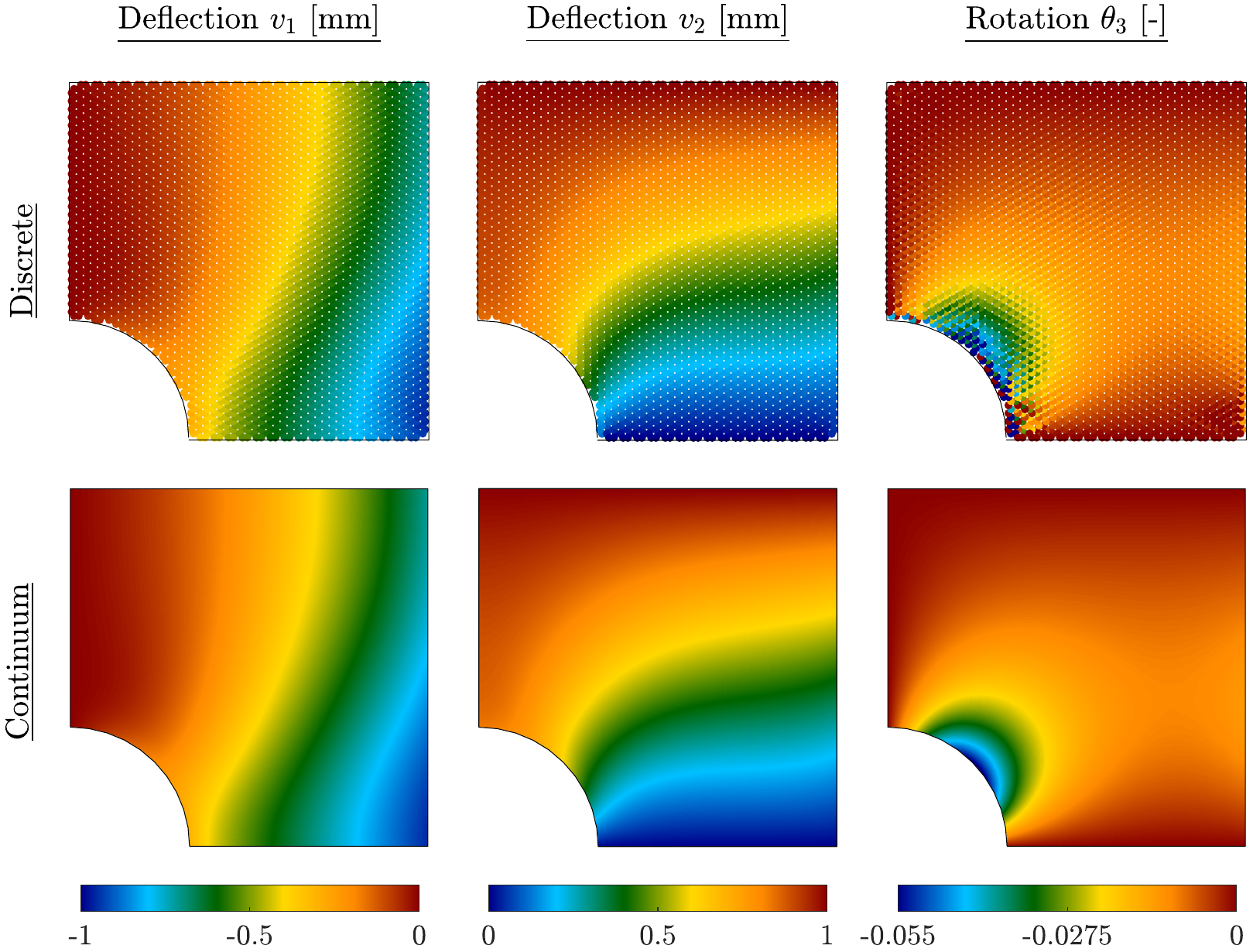}
    \caption{Kinematics of the perforated plate with a honeycomb microstructure. We compare the defections $(v_1,v_2)$ and rotations $\theta_3$ of the discrete metastructure at $1/\epsilon=5$ (top) with the homogenized model (bottom).}
    \label{fig:hip_kin}
\end{figure}

\subsubsection{Perforated plate}\label{sec:hip}

The first problem involves a perforated plate under tension, Fig.~\ref{fig:hip_bvp}. Dirichlet boundary conditions are set as follows: $\theta_3=0$ and $v_1=0$ on the left side; $\theta_3=0$ and $v_2=0$ on the bottom side; and a prescribed vertical deflection $\bar{v}_2=1$ scaled by $1/\epsilon$ on the top side. The honeycomb metamaterial comprises bars with length $L=2$~mm, Fig.~\ref{FIDpWS}, \tB{in-plane thickness $L/10=0.2$~mm}, and unit depth in the out-of-plane direction. We consider a base material with Young's modulus $E=430$~MPa.

Fig.~\ref{fig:hip_kin} shows the resulting kinematics at $1/\epsilon=5$. We observe a notable agreement between the discrete displacement field $u_\epsilon$ from the direct numerical simulation (top row) and the continuum counterpart $u_0$ (bottom row), for deflections and rotations, over the entire domain. Note that the deflections from the discrete model simulation are scaled by $1/\epsilon$. While the vertical displacement field $v_2$ is somewhat expected from the boundary conditions, the accuracy of the lateral displacements $v_1$ and rotations $\theta_3$ resulting from the applied vertical displacement $\bar{v}_2$ is noteworthy. Moreover, $v_1$ reaches a maximum lateral displacement \tB{of about 0.95}, close to the maximum vertical displacement of 1, accurately capturing the significant Poisson effect of the honeycomb microstructure.

\begin{figure}[t!]
    \centering 
    \includegraphics[width=0.85\textwidth]{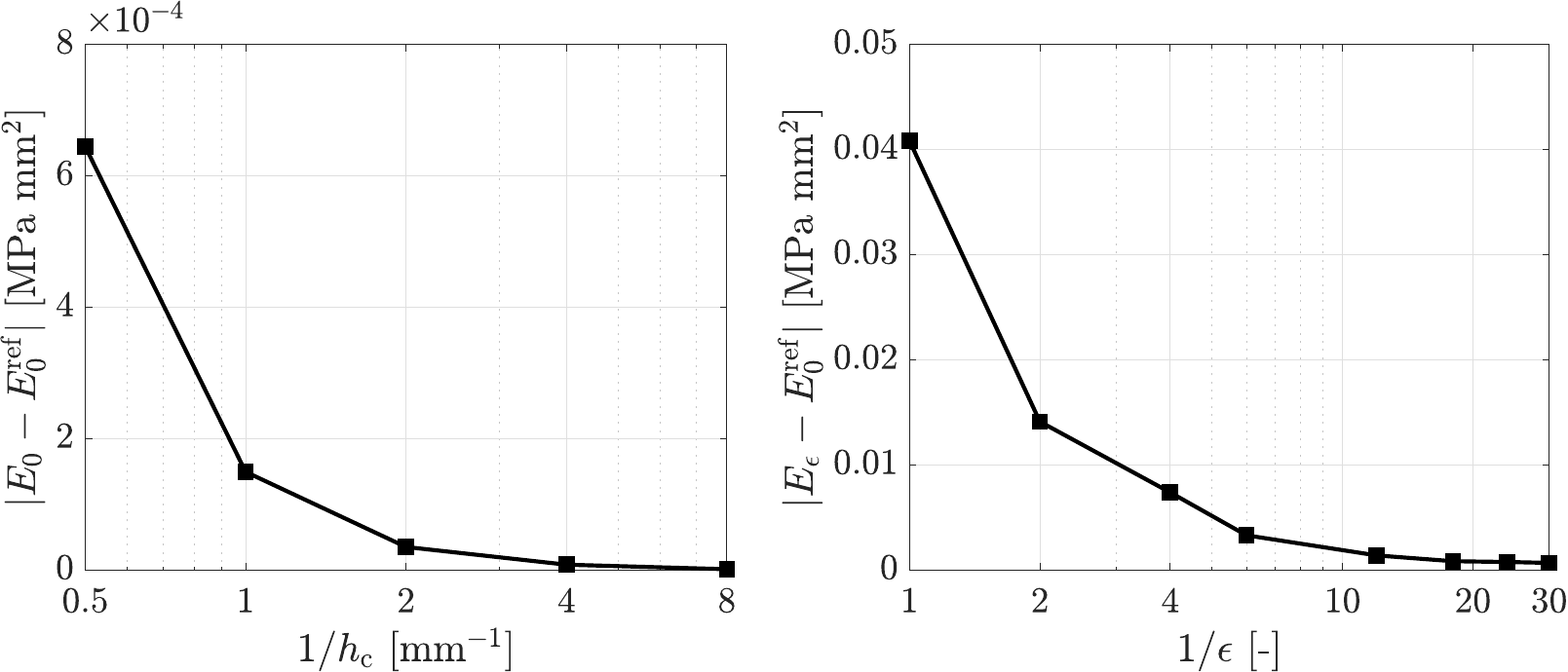}
    \put(-13.8cm,-0cm){\small(a)}
    \put(-6.35cm,-0cm){\small(b)} 
    \vspace{1em}
    \makebox[\textwidth][c]{\hspace{-0.5cm}\includegraphics[width=0.875\textwidth]{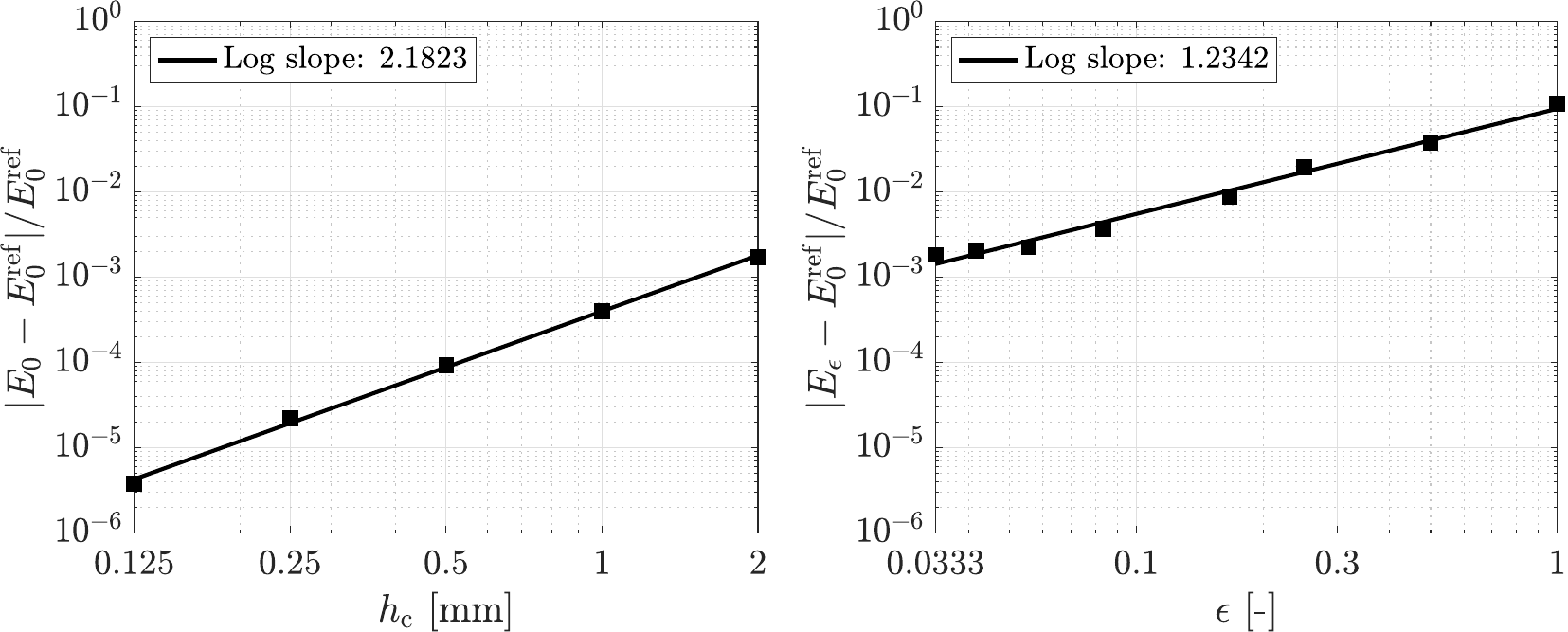}}
    \put(-15.05cm,-0cm){\small(c)}
    \put(-7.6cm,-0cm){\small(d)}   
    \caption{Energy error for the perforated plate. We show (a) absolute and (c) relative errors of the homogenized model as a function of finite element size $h_\mathrm{c}$, and (b) absolute and (d) relative errors between the homogenized model and the discrete metastructure as a function of the scaling factor $\epsilon$.}
    \label{fig:hip_ene_mesh}
\end{figure}

Figures~\ref{fig:hip_ene_mesh}a and~\ref{fig:hip_ene_mesh}c show that the effective energy $E_0$ (Eq.~\eqref{eq:E0hon}), resulting from continuum finite element simulations, converges as the mesh size $h_\mathrm{c}$ decreases. The relative errors remain well below 1\%, with the reference energy $E^\mathrm{ref}_0$ corresponding to the intercept of a least-square fit of the tested mesh sizes, assuming a relation
\begin{equation}
    E_{h_\mathrm{c}} - E_0^\mathrm{ref} \propto h_\mathrm{c}^\alpha,
    \label{eq:ls_fem}
\end{equation}
where $\alpha$ is the convergence rate in logarithmic scale. \tB{The relative error values range from $0.17\%$ at $h_\mathrm{c}=2$~mm} to $0.0004\%$ at $h_\mathrm{c}=0.125$~mm.

\begin{figure}[h!]
    \centering 
    \includegraphics[width=1.0\textwidth]{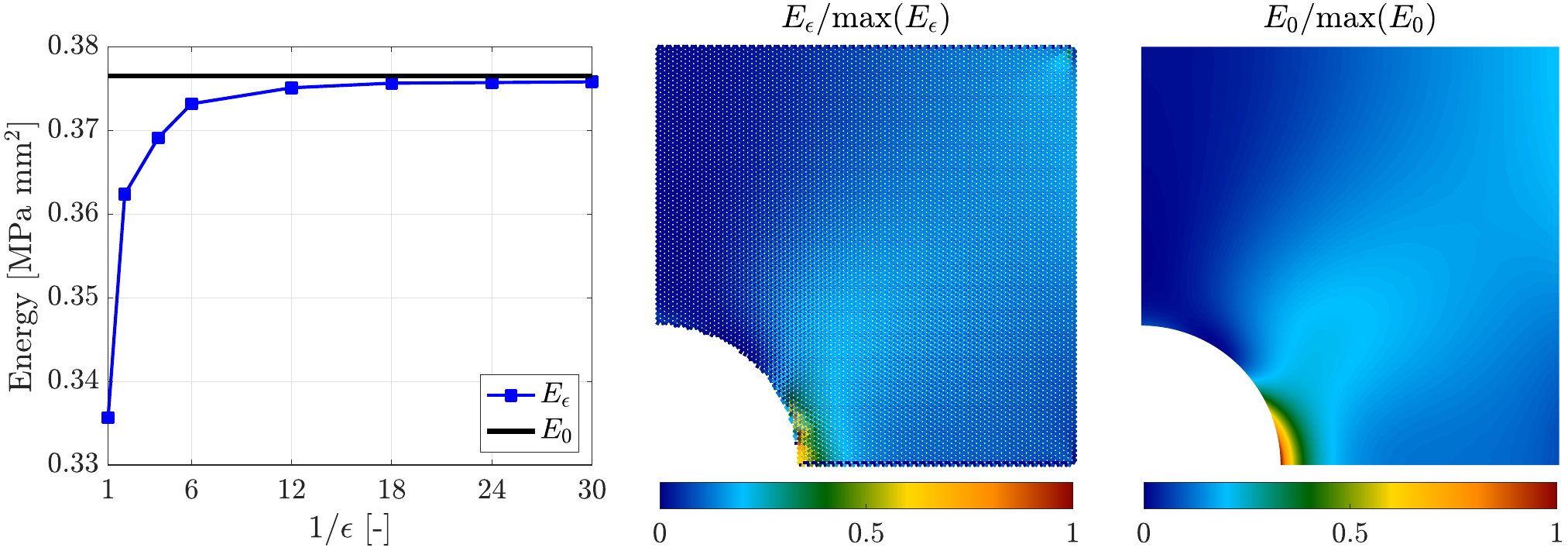}
    \put(-15.15cm,1.12cm){\small(a)}
    \put(-9.6cm,1.12cm){\small(b)}
    \put(-4.5cm,1.12cm){\small(c)}
    \caption{(a) Energy convergence for the perforated plate with a honeycomb microstructure. We show the normalized energy fields of (b) the discrete metastructure and (c) the continuum model.}
    \label{fig:hip_ene}
\end{figure}

Figures~\ref{fig:hip_ene_mesh}b and~\ref{fig:hip_ene_mesh}d further show that the sequence of discrete energies $E_\epsilon$ (Eq.~\eqref{R79mPV}), resulting from direct numerical simulations using Euler-Bernoulli beams, converges to the effective energy $E_0$ as the scaling factor $\epsilon$ decreases. As expected, the relative errors are larger than those in figures~\ref{fig:hip_ene_mesh}a and~\ref{fig:hip_ene_mesh}c, \tB{ranging from $10.8\%$ at $1/\epsilon=1$ to $0.18\%$ at $1/\epsilon=30$}. Nevertheless, a well-defined convergence rate is observed. Fig.~\ref{fig:hip_ene}a further highlights this result, showing the stabilization of the discrete energies with increasing $1/\epsilon$ and their convergence to $E_0$. A relatively stable behavior is observed at $1/\epsilon=18$. Additionally, figures~\ref{fig:hip_ene}b and~\ref{fig:hip_ene}c show a close agreement between the energy density fields over the discrete and continuum domains.

It is worth noting that the continuum simulations entail significant computational savings, even for the rather limited scaling considered herein. In particular, the discrete simulations involved $8\,614$ beam elements and $17\,248$ degrees of freedom for $1/\epsilon=6$, and $213\,966$ beam elements and $428\,004$ degrees of freedom for $1/\epsilon=30$. Conversely, the discretization of the homogenized model is independent of $\epsilon$; it suffices to employ a sufficiently small finite element size such that convergence is achieved, Fig.~\ref{fig:hip_ene_mesh}a. In the present case, the continuum simulations involved $502$ triangular elements and $784$ degrees of freedom for $h_\mathrm{c}=2$~mm, and $30\,573$ elements and $46\,113$ degrees of freedom for $h_\mathrm{c}=0.25$~mm, with the corresponding energies differing by only $0.16\%$. Of course, taking $\epsilon$ to much smaller values, as expected in metamaterials, would render the discrete simulations computationally intractable but would not affect the continuum counterpart.


\subsubsection{Single-edge notched plate}\label{sec:sen}

\begin{figure}[t!]
    \centering 
    \vspace{1em}
    \includegraphics[width=1.0\textwidth]{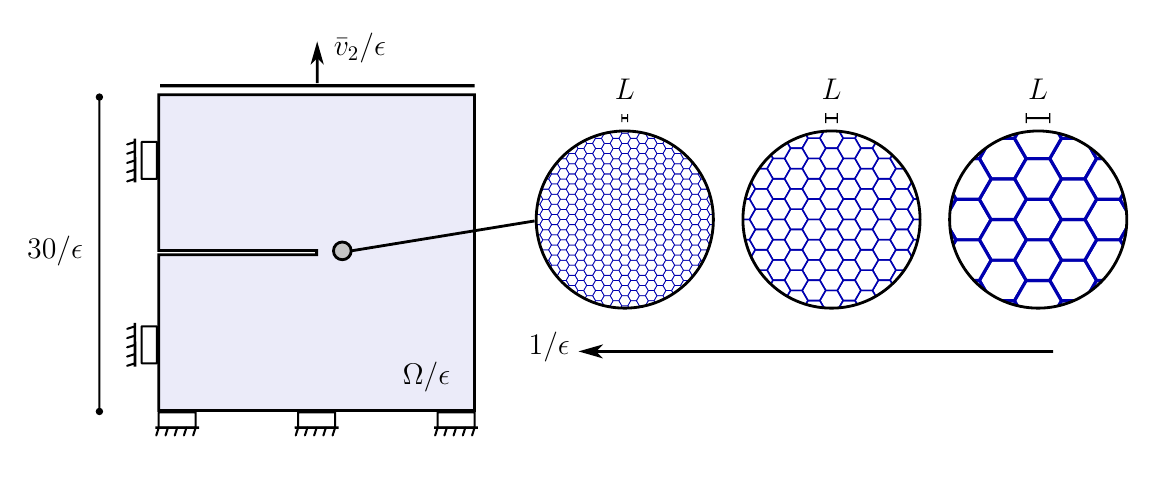}
    \caption{Boundary value problem for the single-edge notched plate with a honeycomb microstructure. The domain is progressively scaled by $1/\epsilon$ with a fixed microstructure size $L$. The scaling is not self-similar due to the fixed, finite crack width. Dimensions in mm.}
    \label{fig:sen_bvp}
\end{figure}

For the sake of illustration, we present a two-dimensional example involving a cracked plate with a honeycomb microstructure, Fig.~\ref{fig:sen_bvp}. The crack is a geometrical notch with a finite width equal to the height of one unit cell, i.e., the minimum possible feature size of the metastructure. Dirichlet boundary conditions are set as follows: $\theta_3=0$ and $v_1=0$ on the left side; $\theta_3=0$, $v_1=0$, and $v_2=0$ on the bottom side; and a prescribed vertical deflection $\bar{v}_2=1$ scaled by $1/\epsilon$ on the top side. The geometrical and material properties of the microstructure are the same as in the previous example.

\begin{figure}[h!]
    \centering 
    \includegraphics[width=0.9\textwidth]{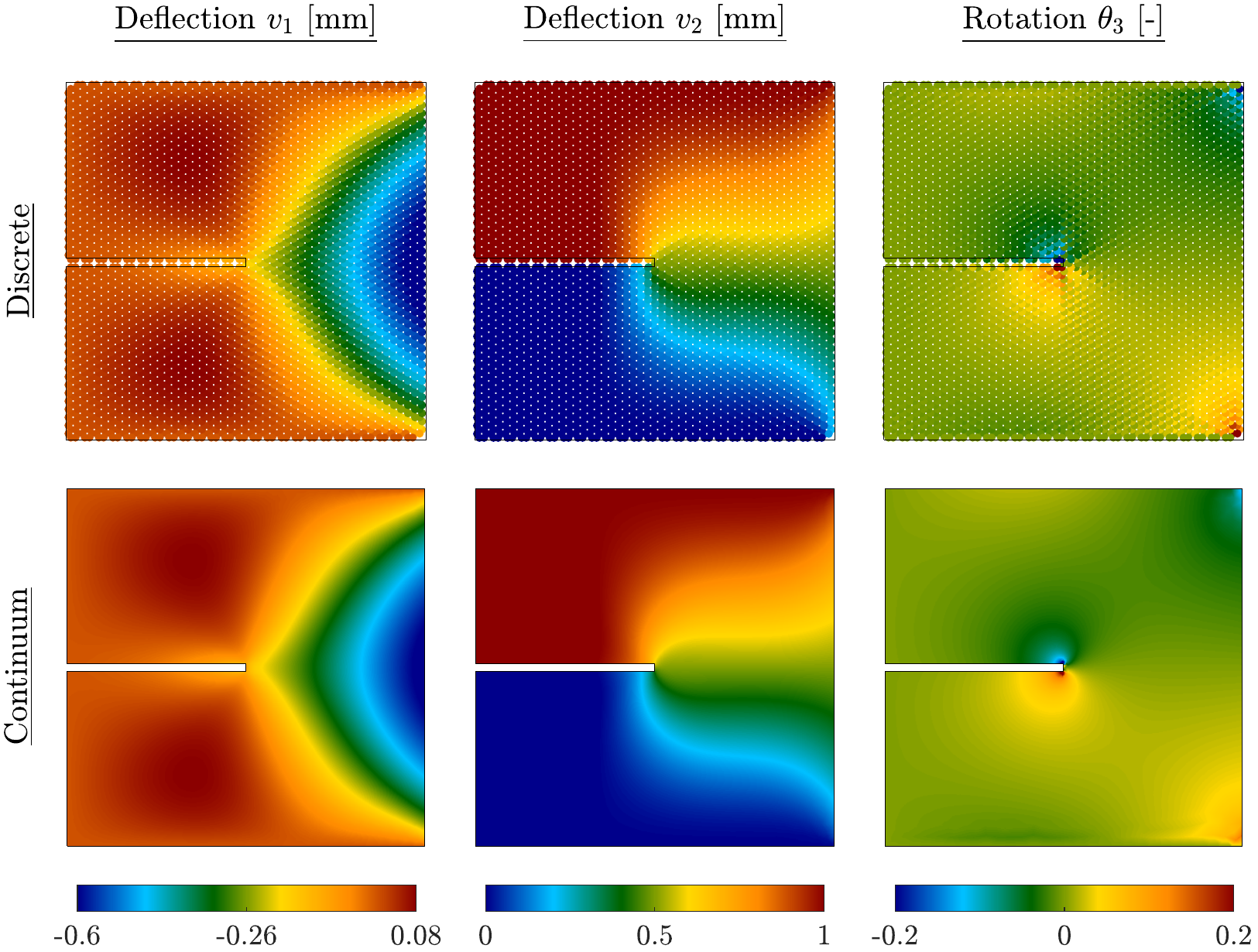}
    \caption{Kinematics of the single-edge notched plate with a honeycomb microstructure. We compare the defections $(v_1,v_2)$ and rotations $\theta_3$ of the discrete metastructure at $1/\epsilon=5$ (top) with the homogenized model (bottom).}
    \label{fig:sen_kin}
\end{figure}

It is important to note that in the present example, the scaling is not self-similar due to the fixed crack width. Consequently, both discrete and continuum models are scaled by $1/\epsilon$, contrary to the previous example, where the continuum finite element simulations were scale-independent. This example thus corresponds to taking successively larger symmetric square cuts around the crack tip of a large metastructure.

\begin{figure}[t!]
    \centering 
    \includegraphics[width=0.85\textwidth]{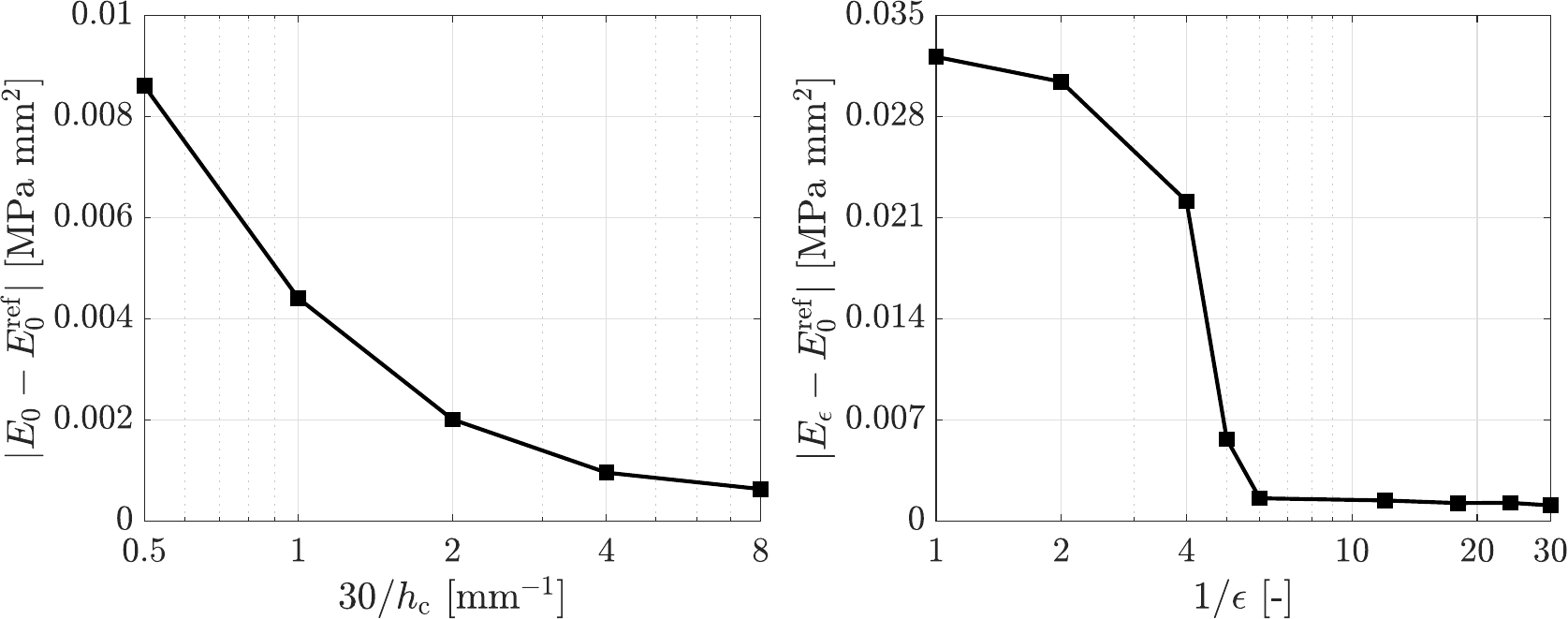}
    \put(-13.8cm,-0cm){\small(a)}
    \put(-6.35cm,-0cm){\small(b)} 
    \vspace{1em}
    \makebox[\textwidth][c]{\hspace{-0.2cm}\includegraphics[width=0.85\textwidth]{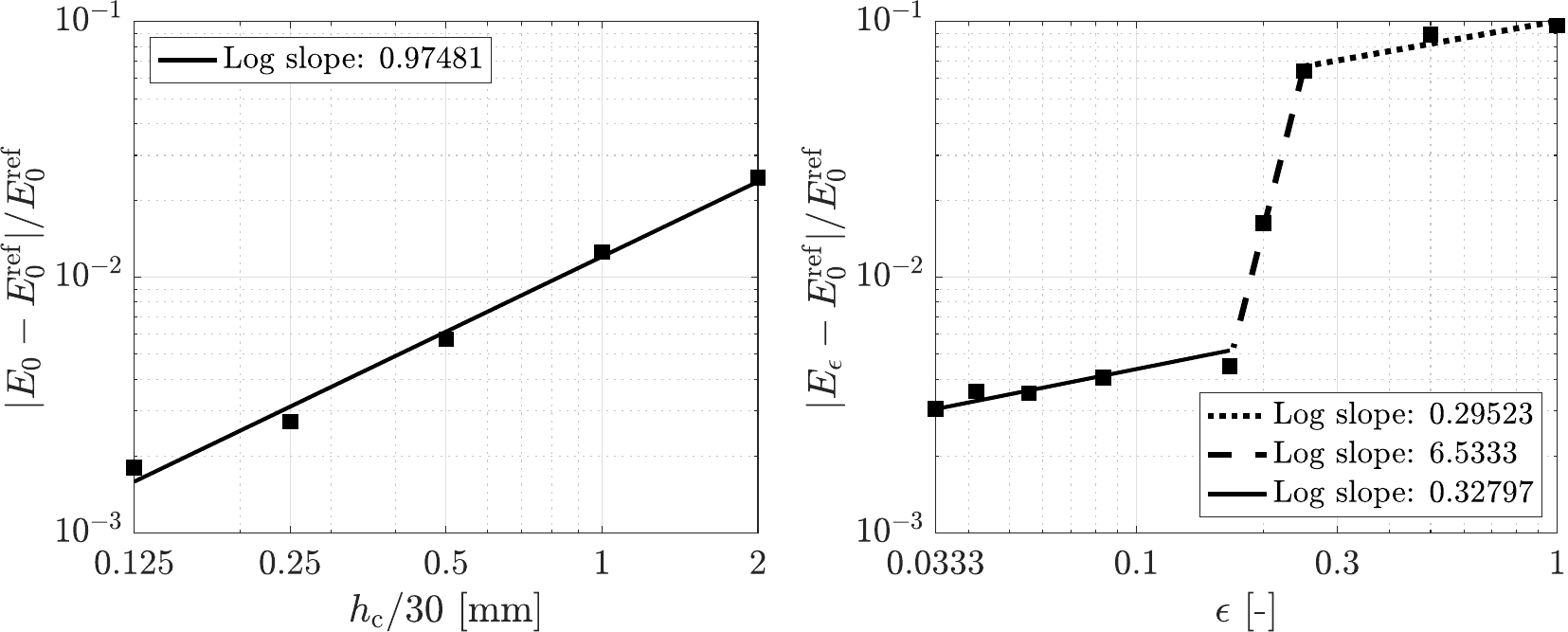}}
    \put(-15.05cm,-0cm){\small(c)}
    \put(-7.6cm,-0cm){\small(d)}      
    \caption{Energy error for the single-edge notched plate. We show (a) absolute and (c) relative errors of the homogenized model as a function of finite element size $h_\mathrm{c}$, and (b) absolute and (d) relative errors between the homogenized model and the discrete metastructure as a function of the scaling factor $\epsilon$.}
    \label{fig:sen_ene_mesh}
\end{figure}

\begin{figure}[h!]
    \centering 
    \includegraphics[width=1.0\textwidth]{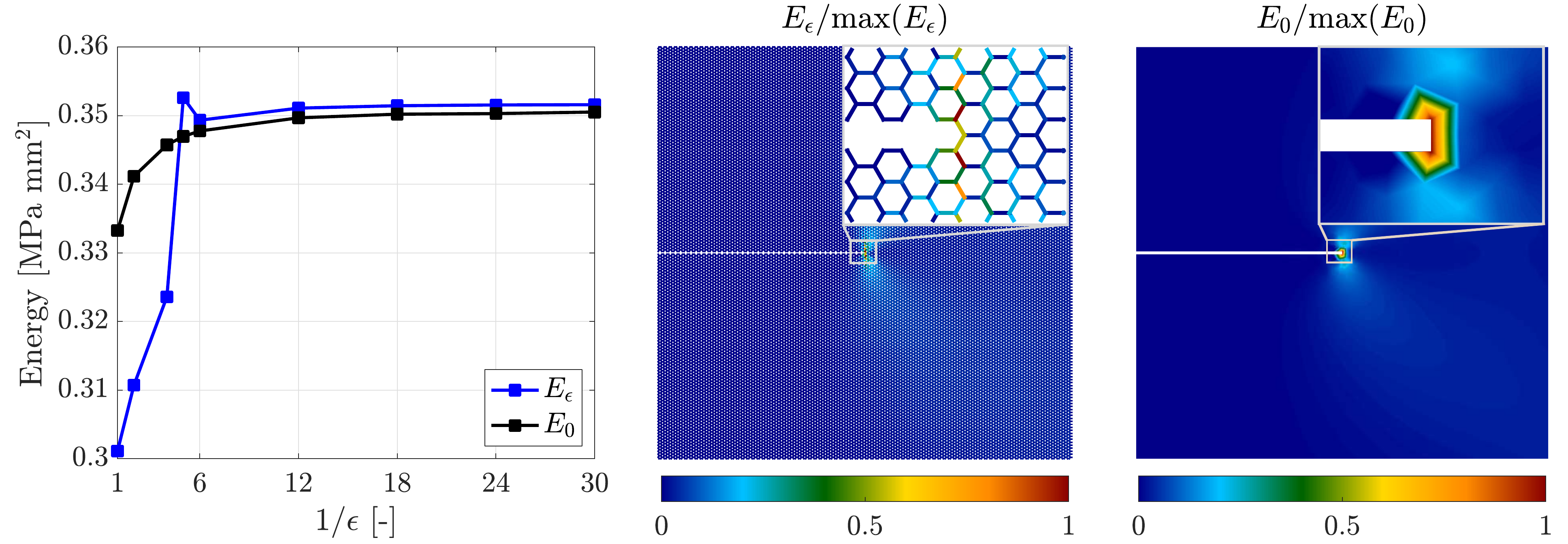}
    \put(-15cm,1.12cm){\small(a)}
    \put(-9.45cm,1.12cm){\color{white}\small(b)}
    \put(-4.37cm,1.12cm){\color{white}\small(c)}            
    \caption{(a) Energy convergence for the single-edge notched plate with a honeycomb microstructure. We show the normalized energy fields of (b) the discrete metastructure and (c) the continuum model. The continuum energy also depends on the scaling factor due to the fixed crack width.}
    \label{fig:sen_ene}
\end{figure}

Fig.~\ref{fig:sen_kin} shows the kinematics at $1/\epsilon=5$. We observe a \tB{ratio between maximum lateral deflections $v_1$ and maximum vertical deflections $v_2$ of about 0.6}. As expected, the rotations $\theta_3$ concentrate around the crack tip, with a clockwise direction above and a counterclockwise direction below. This effect is also observed at the top and bottom right corners. Despite the (finite) crack, we observe an excellent agreement between the response of the discrete metastructure and the homogenized model simulation, considering deflections and rotations across the entire domain. 

As in the previous example, we study the energy convergence numerically. Figures~\ref{fig:sen_ene_mesh}a and~\ref{fig:sen_ene_mesh}c  show the decay of the effective energy $E_0$ resulting from continuum finite element simulations of decreasing mesh size $h_\mathrm{c}$, with $E^\mathrm{ref}_0$ given as in~\eqref{eq:ls_fem}. \tB{The relative errors are higher than in the previous example due to the sharp notch, starting with $2.5\%$ for the coarsest mesh at $h_\mathrm{c}=2$~mm. For smaller mesh sizes, the relative errors remain below $1\%$, reaching $0.18\%$ at $h_\mathrm{c}=0.125$~mm}. Moreover, as before, the errors between the sequence of discrete energies $E_\epsilon$ and the effective energy $E_0$ show comparatively larger values but converge as $\epsilon$ decreases. In particular, \tB{the relative errors drop from $9.64\%$ at $1/\epsilon=1$ to $0.31\%$ at $1/\epsilon=30$}. Note that different convergence regimes are observed in Fig.~\ref{fig:sen_ene_mesh}d, owing to the specimen size approaching the order of magnitude of the fixed crack width at large values of $\epsilon$. Fig.~\ref{fig:sen_ene}a shows the resulting energy convergence curve, where the dependence of $E_0$ on $\epsilon$ reflects the effect of the fixed crack width. The energy density fields over the discrete and continuum domains in figures~\ref{fig:sen_ene}b and~\ref{fig:sen_ene}c show a close agreement, concentrating around the crack tip.

In this example, the discrete simulations involved $9\,417$ beam elements and $18\,715$ degrees of freedom for $1/\epsilon=6$, and $234\,293$ beam elements and $467\,987$ degrees of freedom for $1/\epsilon=30$. On the other hand, the continuum simulations involved $706$ triangular elements and $1\,124$ degrees of freedom for $h_\mathrm{c}=2$~mm, and $39\,959$ elements and $60\,477$ degrees of freedom for $h_\mathrm{c}=0.25$~mm. We emphasize that the computational burden of simulating smaller microstructures (or, equivalently, larger structures) becomes intractable for discrete simulations, while the computational requirements of the continuum model are unaffected.


\begin{figure}[b!]
    \centering 
    \vspace{1em}
    \includegraphics[width=0.95\textwidth]{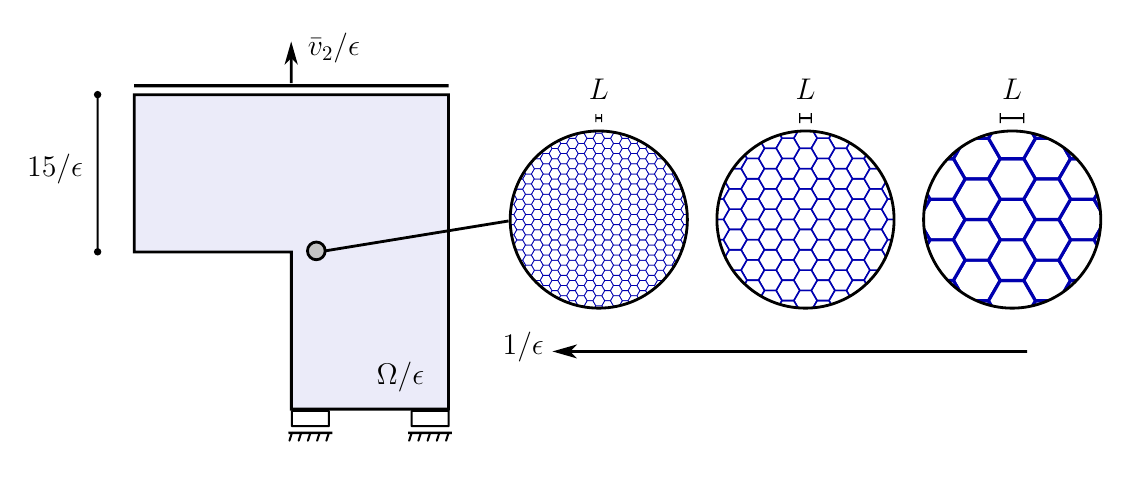}
    \caption{Boundary value problem for the L-shaped plate with a honeycomb microstructure. The domain is progressively scaled by $1/\epsilon$ with a fixed microstructure size $L$. Dimensions in mm.}
    \label{fig:L2D_bvp}
\end{figure}

\subsubsection{L-shaped plate}

We consider another two-dimensional problem consisting of an L-shaped specimen with a honeycomb microstructure under tensile loading, Fig.~\ref{fig:L2D_bvp}. Dirichlet boundary conditions are set as $\theta_3=0$ and $v_2=0$ on the bottom side and a prescribed vertical deflection $\bar{v}_2=1$ scaled by $1/\epsilon$ on the top side. The geometrical and material properties of the microstructure are the same as in the previous example.

\begin{figure}[t!]
    \centering 
    \includegraphics[width=0.9\textwidth]{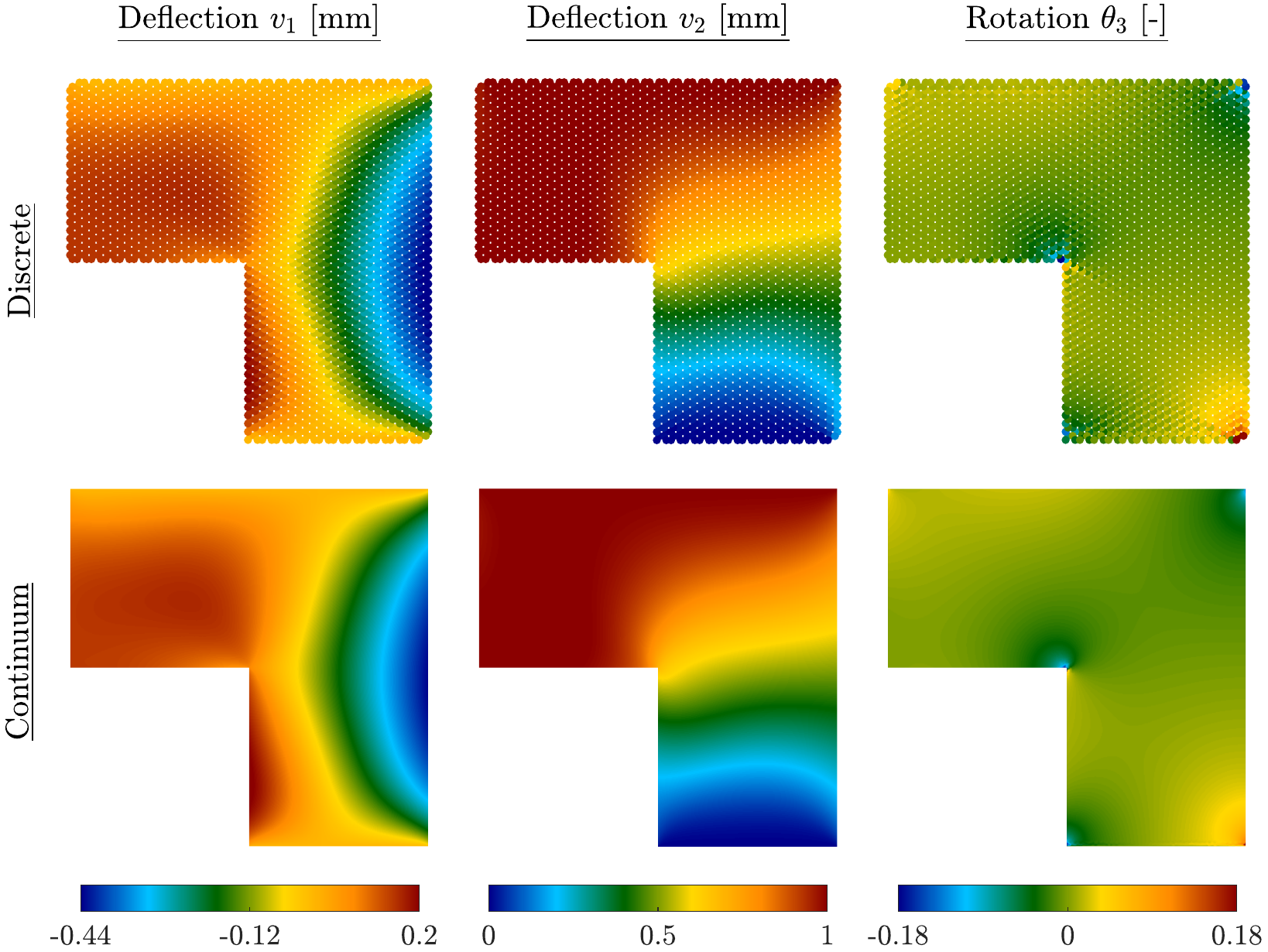}
    \caption{Kinematics of the L-shaped plate with a honeycomb microstructure. We compare the defections $(v_1,v_2)$ and rotations $\theta_3$ of the discrete metastructure at $1/\epsilon=5$ (top) with the homogenized model (bottom).}
    \label{fig:L2D_kin}
\end{figure}

Fig.~\ref{fig:L2D_kin} shows the kinematics at $1/\epsilon=5$. The direct numerical simulations show a ratio between maximum lateral deflections $v_1$ and maximum vertical deflections $v_2$ of about 0.44. Moreover, the rotations $\theta_3$ strongly concentrate at the corners. As in the previous example, the homogenized model accurately captures this behavior, showing a notable quantitative agreement over the entire domain for the deflection components and rotations, despite the effects of the inner corner.

\begin{figure}[t!]
    \centering 
    \includegraphics[width=0.85\textwidth]{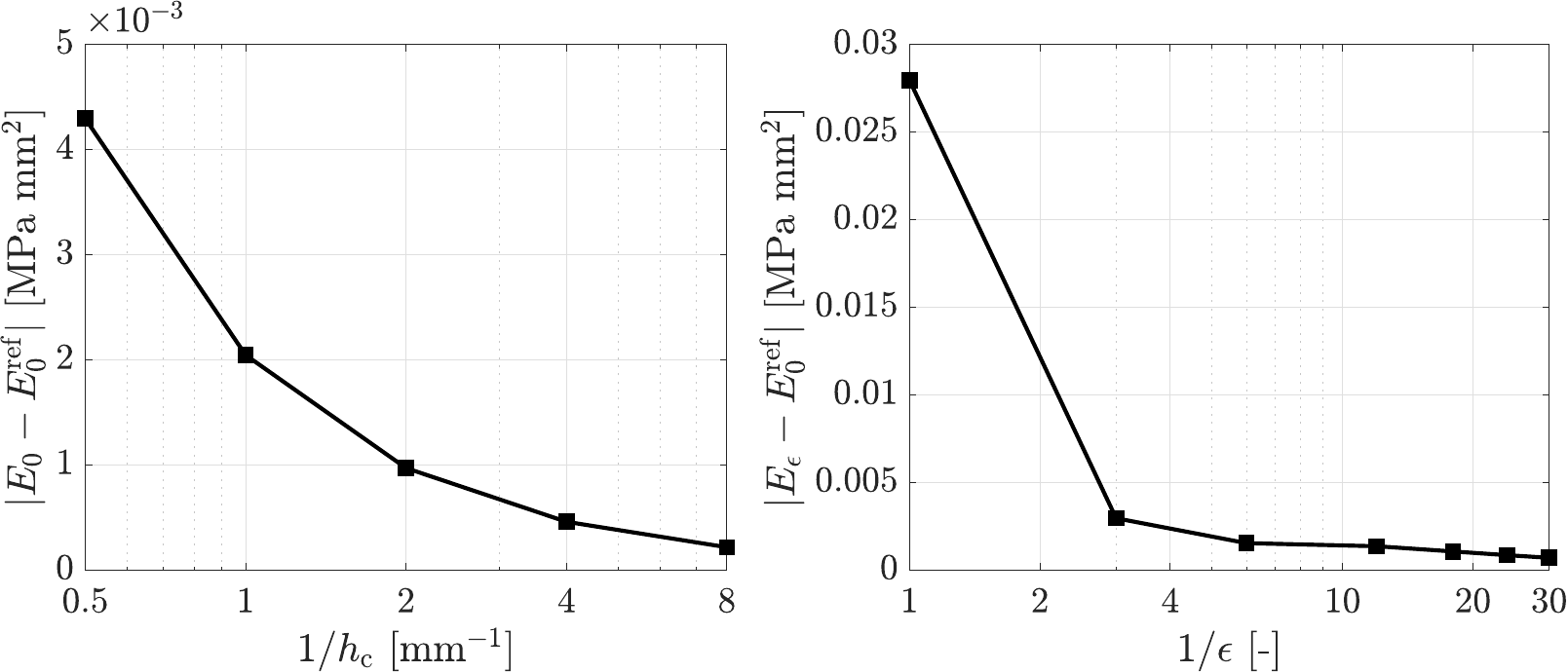}
    \put(-13.8cm,-0cm){\small(a)}
    \put(-6.35cm,-0cm){\small(b)} 
    \vspace{1em}
    \makebox[\textwidth][c]{\hspace{-0.5cm}\includegraphics[width=0.85\textwidth]{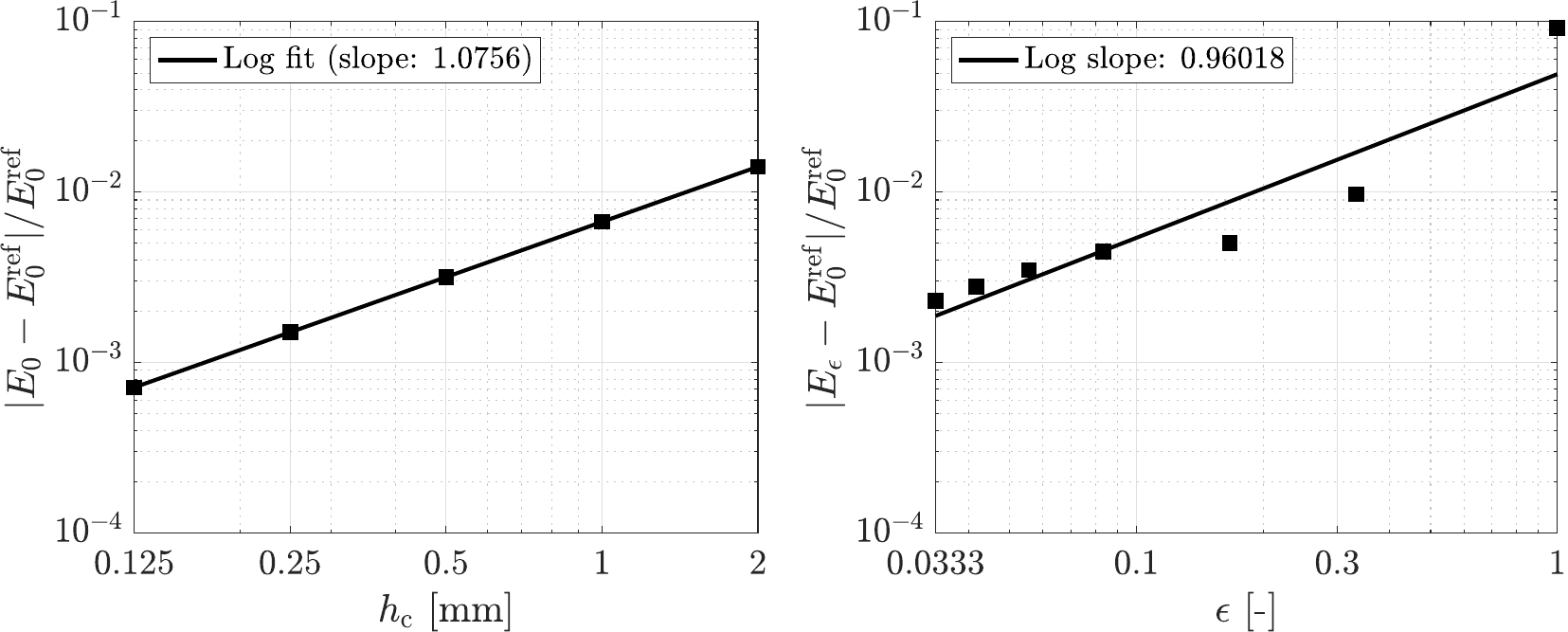}}
    \put(-15.05cm,-0cm){\small(c)}
    \put(-7.6cm,-0cm){\small(d)} 
    \caption{Energy error for the L-shaped plate. We show (a) absolute and (c) relative errors of the homogenized model as a function of finite element size $h_\mathrm{c}$, and (b) absolute and (d) relative errors between the homogenized model and the discrete metastructure as a function of the scaling factor $\epsilon$.}
    \label{fig:L2D_ene_mesh}
\end{figure}

\begin{figure}[h!]
    \centering 
    \includegraphics[width=1.0\textwidth]{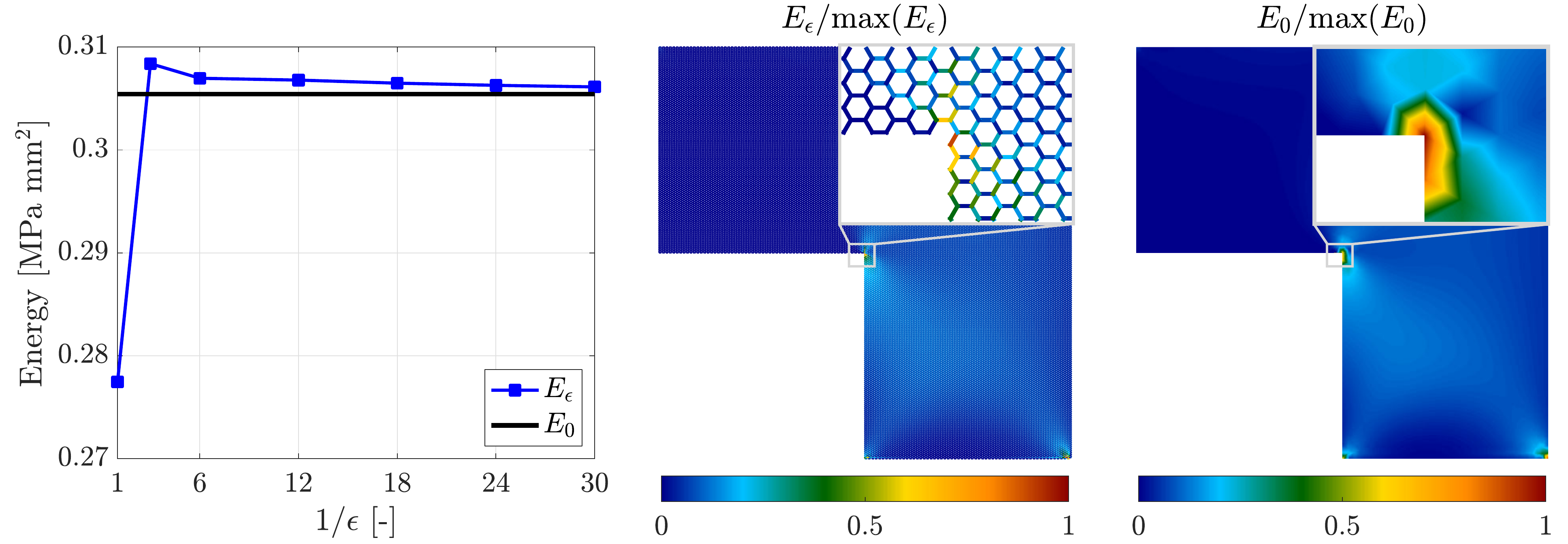}
    \put(-15.075cm,1.12cm){\small(a)}
    \put(-9.6cm,1.12cm){\small(b)}
    \put(-4.5cm,1.12cm){\small(c)}
    \caption{(a) Energy convergence for the L-shaped plate with a honeycomb microstructure. We show the normalized energy fields of (b) the discrete metastructure and (c) the continuum model.}
    \label{fig:L2D_ene}
\end{figure}

We proceed to assess the energy convergence numerically. Figures~\ref{fig:L2D_ene_mesh}a and~\ref{fig:L2D_ene_mesh}c  show the decay of the effective energy $E_0$ resulting from continuum finite element simulations of decreasing mesh size $h_\mathrm{c}$, with $E^\mathrm{ref}_0$ given as in~\eqref{eq:ls_fem}. The errors, modulated by the sharp corner, are higher than in the perforated plate of Sec.~\ref{sec:hip}, but smaller than in the single-edge notched sample of Sec.~\ref{sec:sen}. We observe a \tB{relative error of $1.4\%$ for the coarsest mesh at $h_\mathrm{c}=2$~mm, which drops to $0.07\%$ at $h_\mathrm{c}=0.125$~mm}. On the other hand, the errors between the sequence of discrete energies $E_\epsilon$ and the effective energy $E_0$ converge as $\epsilon$ decreases. The \tB{relative errors drop from $9.16\%$ at $1/\epsilon=1$ to $0.23\%$ at $1/\epsilon=30$}, showing a well-defined convergence rate. Fig.~\ref{fig:L2D_ene} presents the resulting energy convergence curve and the excellent agreement of the energy density across the entire domain.

In this example, the discrete simulations involved $7\,152$ beam elements and $14\,409$ degrees of freedom for $1/\epsilon=6$, and $176\,168$ beam elements and $352\,857$ degrees of freedom for $1/\epsilon=30$. On the other hand, the continuum simulations involved $436$ triangular elements and $699$ degrees of freedom for $h_\mathrm{c}=2$~mm, and $25\,256$ elements and $38\,243$ degrees of freedom for $h_\mathrm{c}=0.25$~mm. We emphasize that the computational burden of simulating smaller microstructures (or, equivalently, larger structures) becomes intractable for discrete simulations, while the computational requirements of the continuum model are unaffected. Of course, the computational savings are more significant in three-dimensional settings, as discussed in the following examples.

\subsection{Three-dimensional octet-truss metastructures}

We now assess the limiting continuum model of three-dimensional octet-truss metamaterials, derived in Example~\ref{ex_octet}, for different boundary value problems.

\begin{figure}[t!]
    \centering 
    \vspace{1em}
    \includegraphics[width=1.0\textwidth]{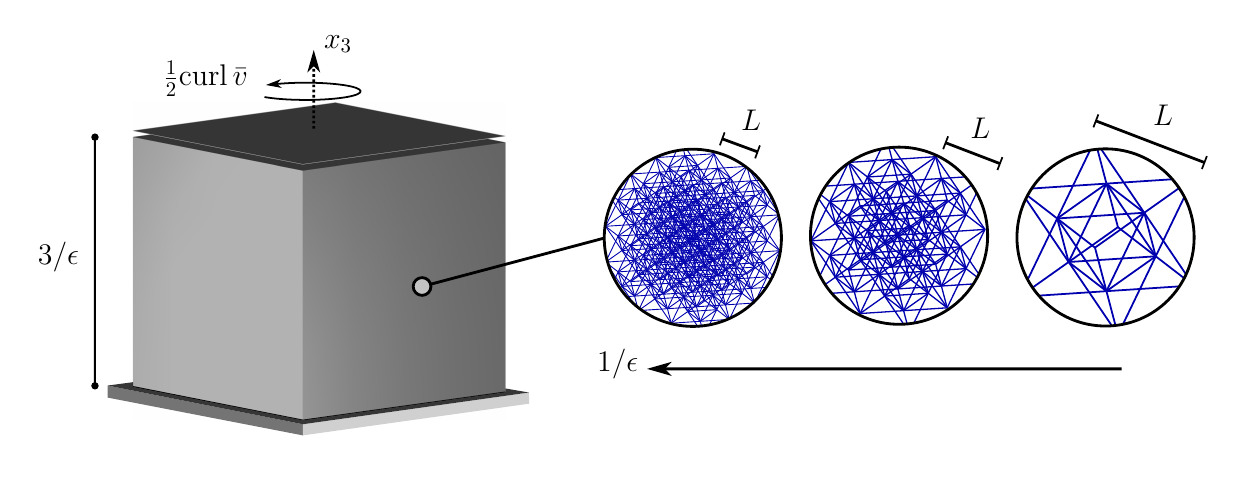}
    \caption{Boundary value problem for the octet-truss cube under torsion. The domain is progressively scaled by $1/\epsilon$ with a fixed microstructure size $L$. Dimensions in mm.}
    \label{fig:rotcube_bvp}
\end{figure}

\begin{figure}[h!]
    \centering 
    \includegraphics[width=0.82\textwidth]{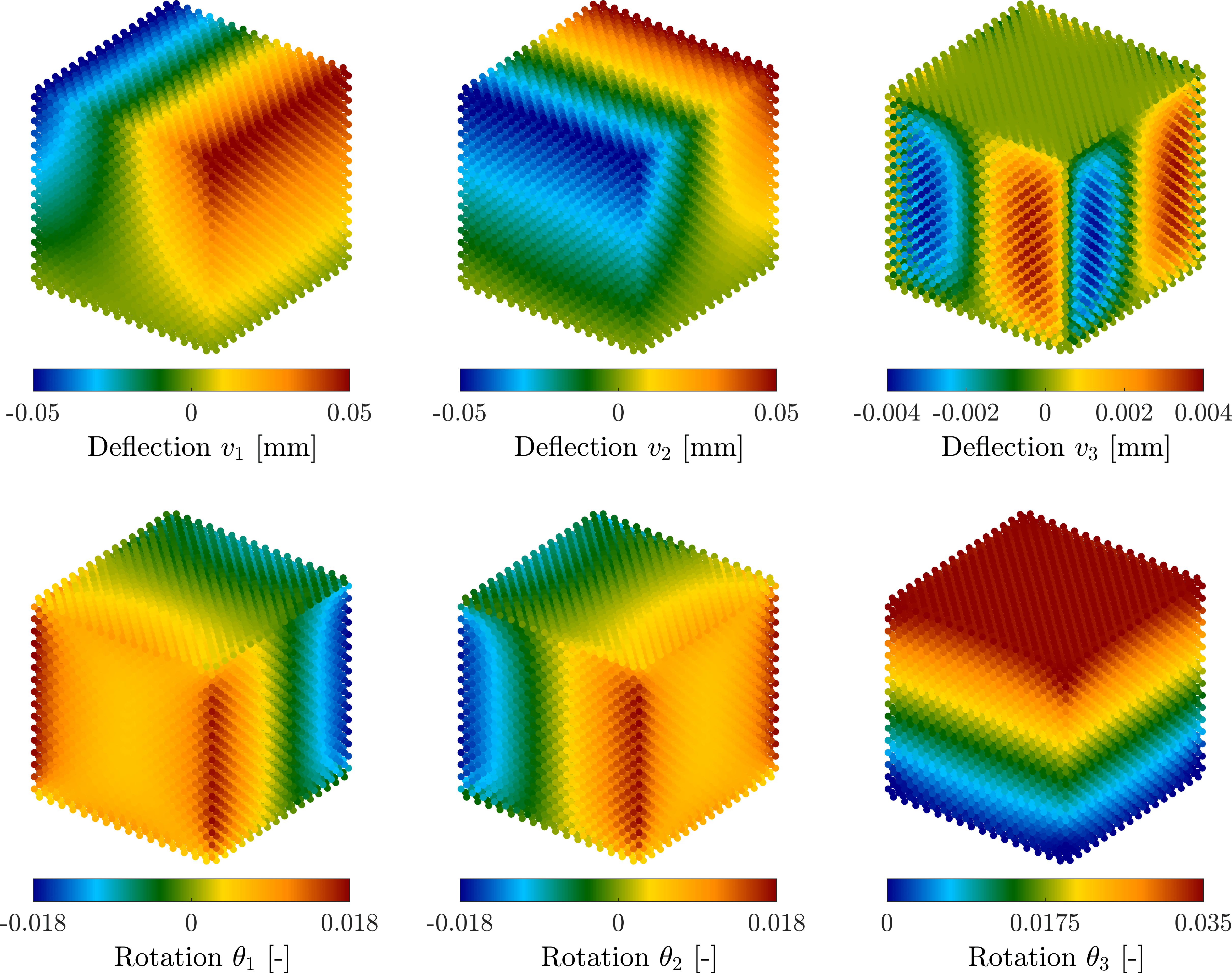}
    \caption{Kinematics of the octet-truss cube under torsion resulting from the direct numerical simulation of the discrete metastructure at $1/\epsilon=6$.}
    \label{fig:rotcube_kin_disc}
\end{figure}

\subsubsection{Cube torsion}

\begin{figure}[t!]
    \centering 
    \includegraphics[width=0.82\textwidth]{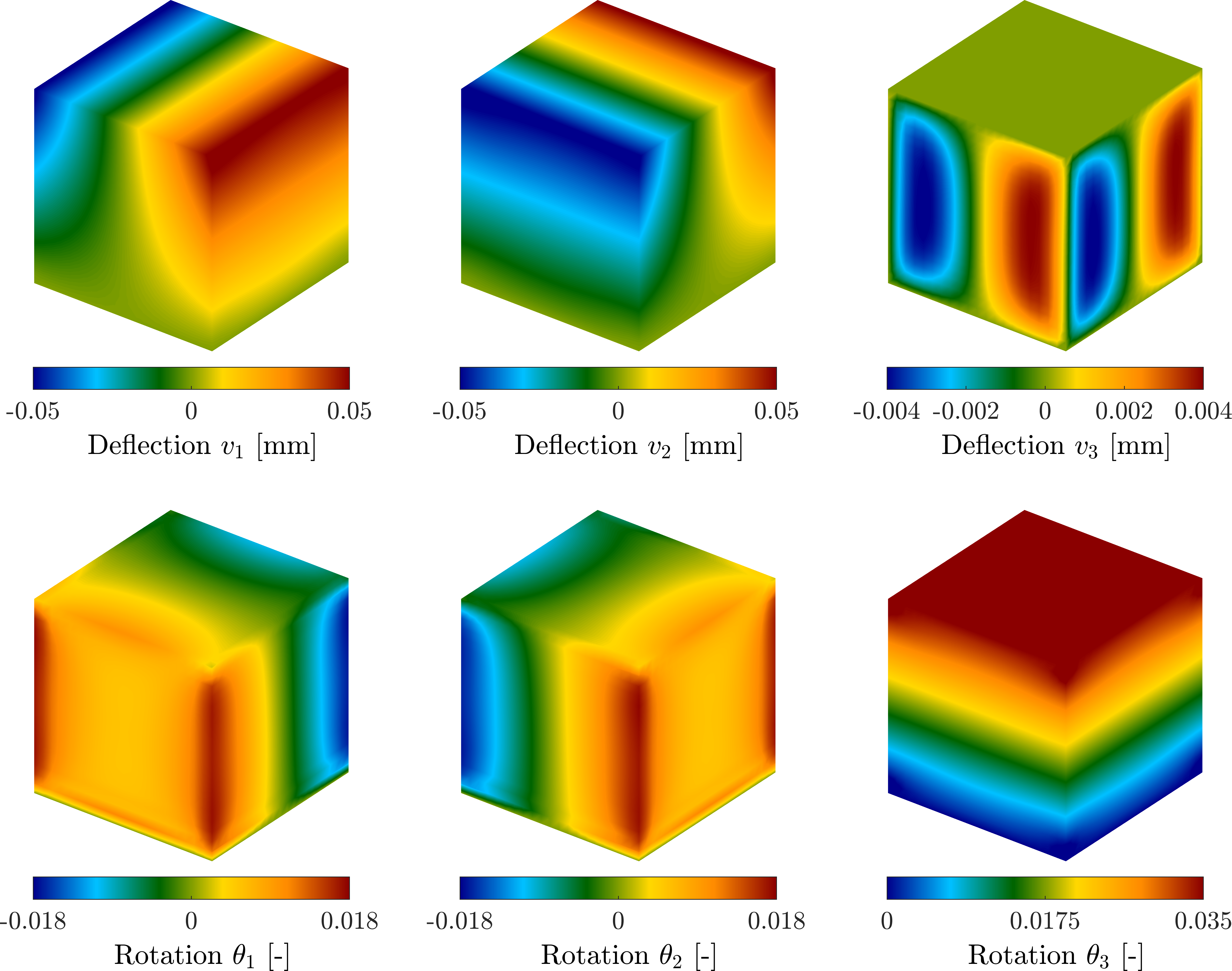}
    \caption{Kinematics of the octet-truss cube under torsion resulting from the homogenized model simulation.}
    \label{fig:rotcube_kin_cont}
\end{figure}

\begin{figure}[h!]
    \centering 
    \includegraphics[width=1\textwidth]{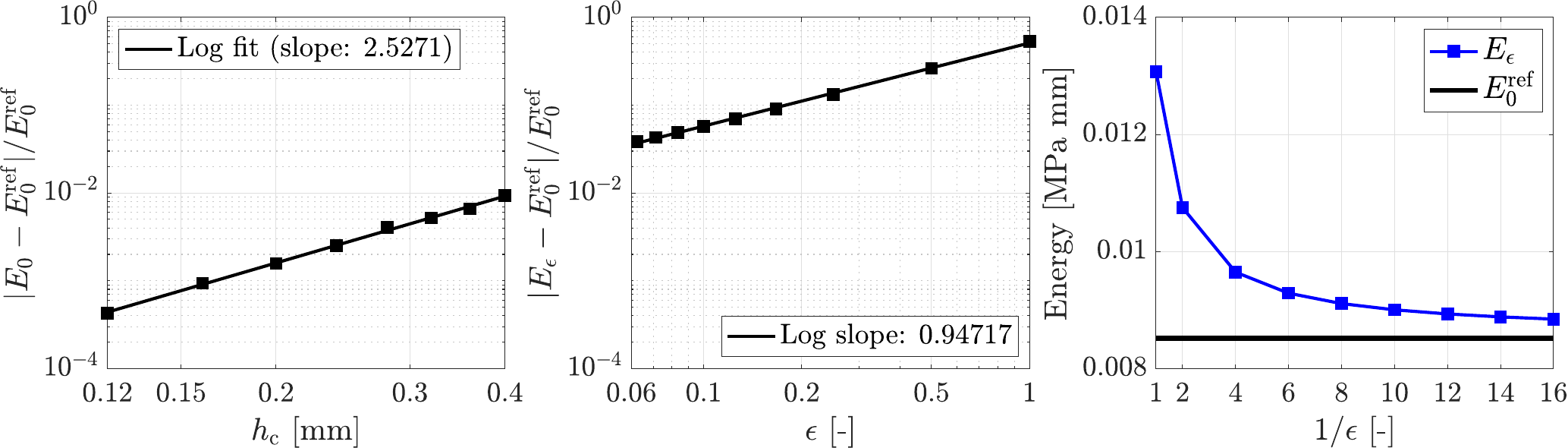}
    \put(-16cm,0cm){\small(a)}
    \put(-10.5cm,0cm){\small(b)}
    \put(-4.9cm,0cm){\small(c)}    
    \caption{Energy convergence and relative errors for the octet-truss cube under torsion.}
    \label{fig:rotcube_ene_relene_mesh}
\end{figure}

The first three-dimensional example considers a cube with an octet-truss microstructure under torsional loading, Fig.~\ref{fig:rotcube_bvp}. Dirichlet boundary conditions are set with $v_1=v_2=v_3=0$ on the bottom face and a compatible rotation $\bar{\vartheta}=2\pi/180$ on the top face, imposed by setting
\begin{equation}
\begin{aligned}
v_1(x_1,x_2,3/\epsilon)=\bar{v}_1=x_1\cos{\bar{\vartheta}}-x_2\sin{\bar{\vartheta}},\\
v_2(x_1,x_2,3/\epsilon)=\bar{v}_2=x_1\sin{\bar{\vartheta}}+x_2\cos{\bar{\vartheta}}.
\end{aligned}
\end{equation}
On the other hand, the incompatible rotations $\theta$ are left free at the boundary. The octet-truss metamaterial has a unit cell length $L=0.75$~mm, Fig.~\ref{S2rFkk}. The bars have a circular cross-section with a diameter of \tB{$0.065$~mm}. We consider a base material with Young's modulus $E=430$~MPa. These parameters correspond to the octet-truss structures studied in~\cite{Shaikeea2022}.

Figure~\ref{fig:rotcube_kin_disc} shows the deflection components $(v_1,v_2,v_3)$ and rotation components $(\theta_1,\theta_2,\theta_3)$ of the discrete metastructure scaled at $1/\epsilon=6$. Note that $\theta_3$ is close to $2\pi/180$ on the top face. As such, the rotations are virtually compatible and the axial term in the effective energy~\eqref{GQ4mav} is dominant, as expected from the loading and boundary conditions. Figure~\ref{fig:rotcube_kin_cont} shows the results of the homogenized model in terms of the limiting energy~\eqref{GQ4mav}. We observe a remarkable agreement with the direct numerical simulation for all displacement and rotation components across the entire domain.

As in the two-dimensional examples of Sec.~\ref{sec:comp2D}, we assess the numerical convergence of the three-dimensional octet-truss metastructures. Fig.~\ref{fig:rotcube_ene_relene_mesh}a highlights the reliability of the continuum finite element simulations, exhibiting asymptotic convergence as the mesh size $h_\mathrm{c}$ decreases, with $E^\mathrm{ref}_0$ given in~\eqref{eq:ls_fem}. The relative errors drop from $0.93$\% at $h_\mathrm{c}=0.4$~mm to $0.04$\% at $h_\mathrm{c}=0.12$~mm. On the other hand, figures~\ref{fig:rotcube_ene_relene_mesh}b and~\ref{fig:rotcube_ene_relene_mesh}c show the convergence of the sequence of discrete energies~\eqref{R79mPV} to the continuum limit~\eqref{GQ4mav} as $\epsilon$ decreases. The relative errors range from $53.53$\% at $\epsilon=1$ to $3.85$\% at $1/\epsilon=16$, showing a well-defined convergence rate.

The errors are higher than in the two-dimensional examples because the three-dimensional discrete metastructures analyzed herein are coarser (cf. $1/\epsilon=30$ in the two-dimensional tests), as finer metastructures imply a significantly large number of degrees of freedom. Indeed, the computational savings in the present three-dimensional simulations are noteworthy: the largest discrete model ($1/\epsilon=16$) involved $6\,340\,608$ elements and $6\,406\,784$ degrees of freedom, while the finest continuum model involved $13\,873$ tetrahedral elements and $69\,900$ degrees of freedom, independent of $\epsilon$.

\begin{figure}[b!]
    \centering 
    \includegraphics[width=1.0\textwidth]{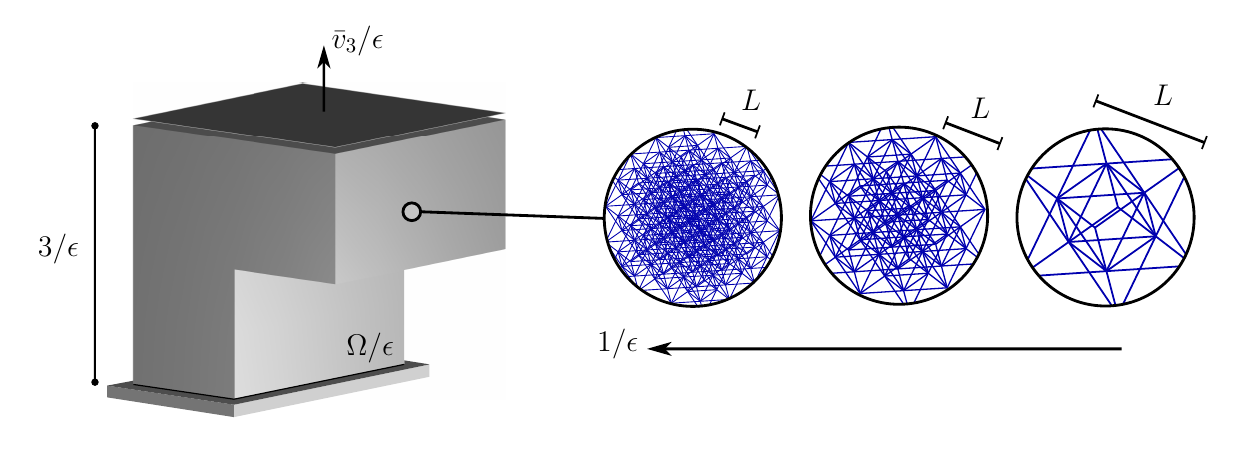}
    \caption{Boundary value problem for the L-shaped specimen with an octet-truss microstructure. The domain is progressively scaled by $1/\epsilon$ with a fixed microstructure size $L$. Dimensions in mm.}
    \label{fig:L_bvp}
\end{figure}

\begin{figure}[t!]
    \centering 
    \includegraphics[width=0.77\textwidth]{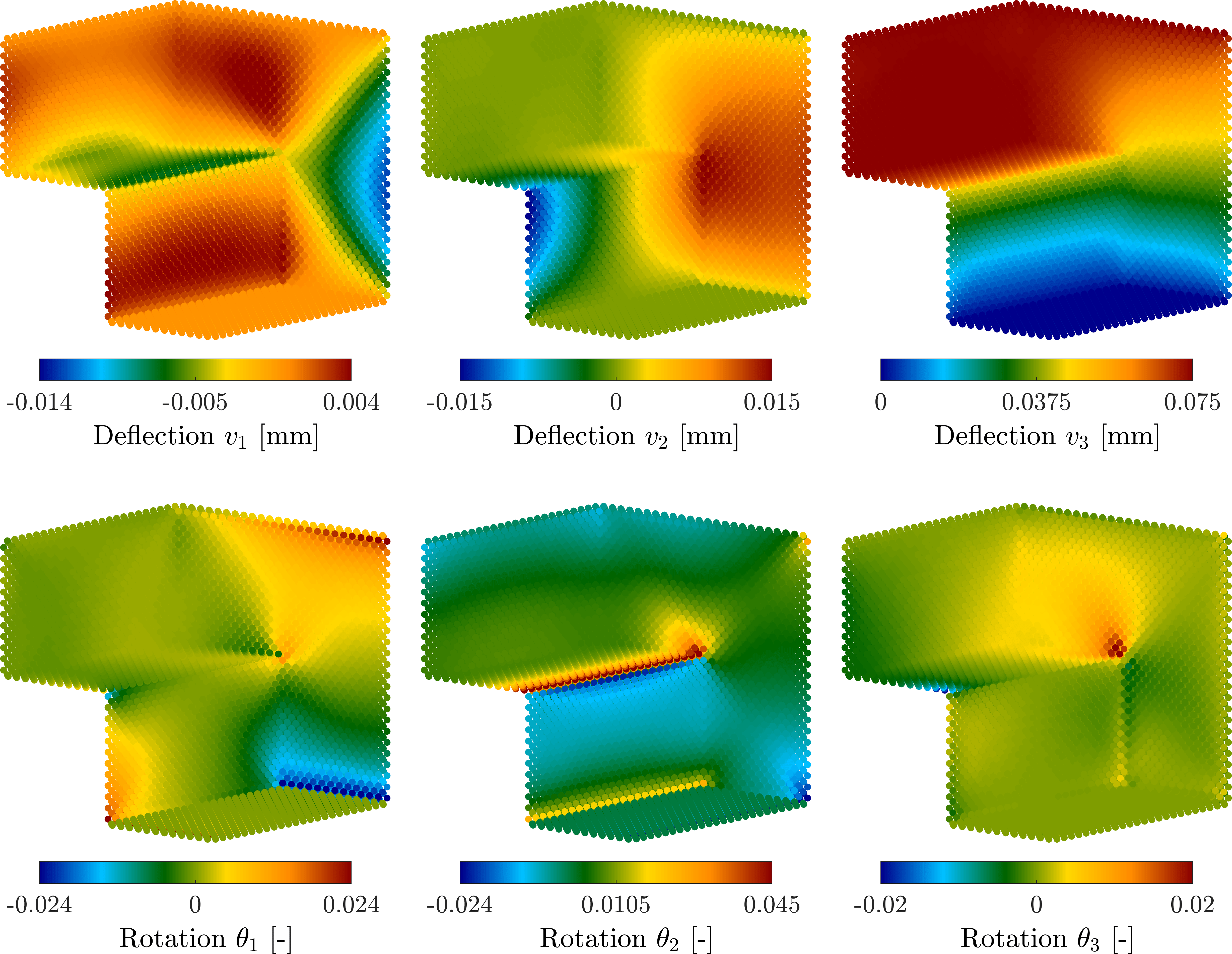}
    \caption{Kinematics of the octet-truss L-shaped specimen resulting from the direct numerical simulation of the discrete metastructure at $1/\epsilon=6$.}
    \label{fig:L_kin_disc}
\end{figure}

\begin{figure}[h!]
    \centering 
    \includegraphics[width=0.77\textwidth]{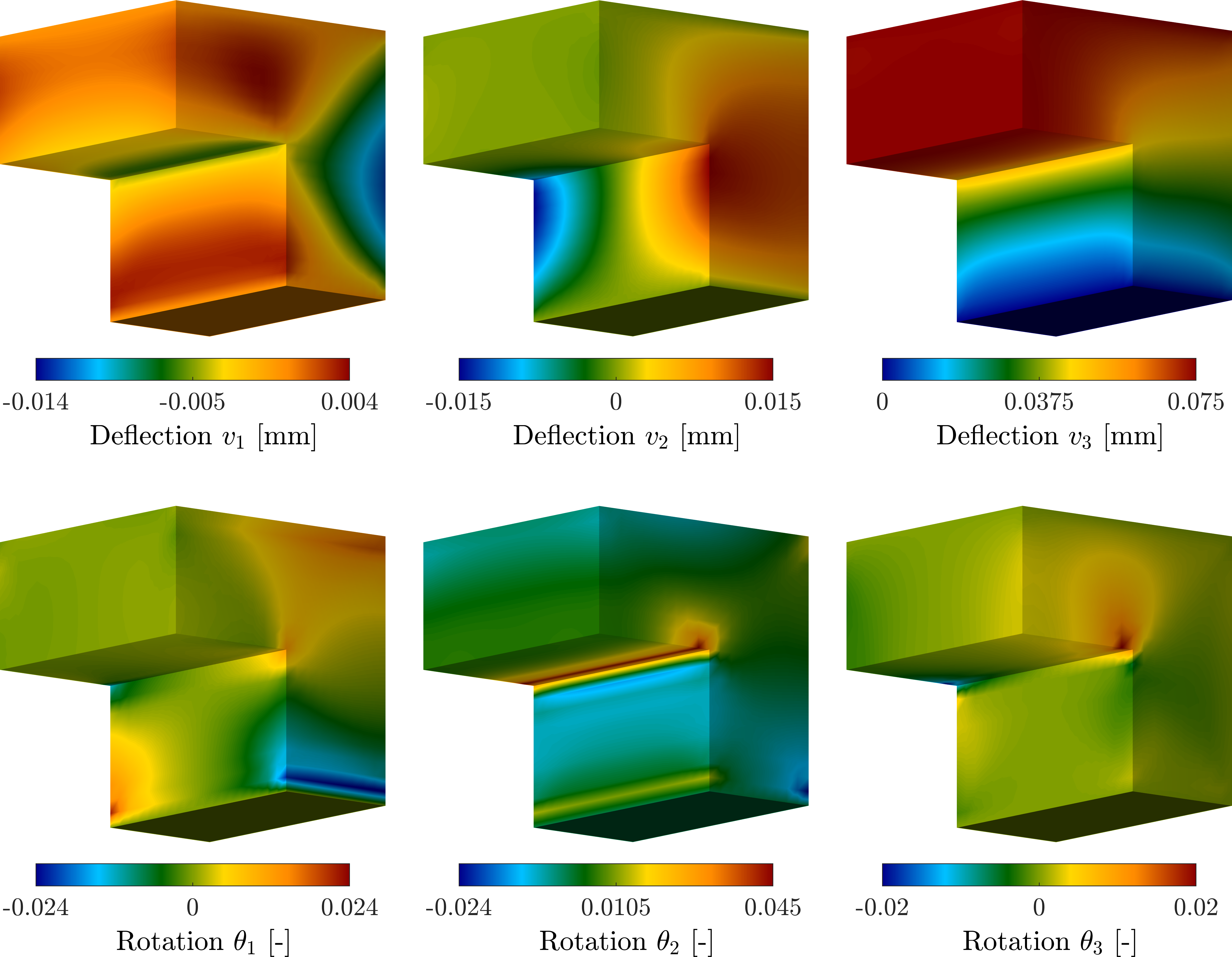}
    \caption{Kinematics of the octet-truss L-shaped specimen resulting from the homogenized model simulation.}
    \label{fig:L_kin_cont}
\end{figure}

\subsubsection{L-shaped specimen}

The final example revisits the L-shaped specimen in a three-dimensional setting, with an octet-truss microstructure, Fig.~\ref{fig:L_bvp}. Dirichlet boundary conditions are set as $\theta_1=\theta_2=\theta_3=0$ and $v_1=v_2=v_3=0$ on the bottom face, and a prescribed vertical deflection $\bar{v}_3=1$ scaled by $1/\epsilon$ on the top face. The geometry of the bars and the base material properties are the same as in the previous example.

Figure~\ref{fig:L_kin_disc} shows the deflection components $(v_1,v_2,v_3)$ and rotation components $(\theta_1,\theta_2,\theta_3)$ of the discrete metastructure scaled at $1/\epsilon=6$. As expected, the rotations concentrate at the inner corner and close to the boundaries. Figure~\ref{fig:L_kin_cont} shows an excellent agreement between the discrete metastructure and the kinematics predicted by the homogenized model~\eqref{GQ4mav} for all displacement and rotation components.

\begin{figure}[t!]
    \centering 
    \includegraphics[width=1\textwidth]{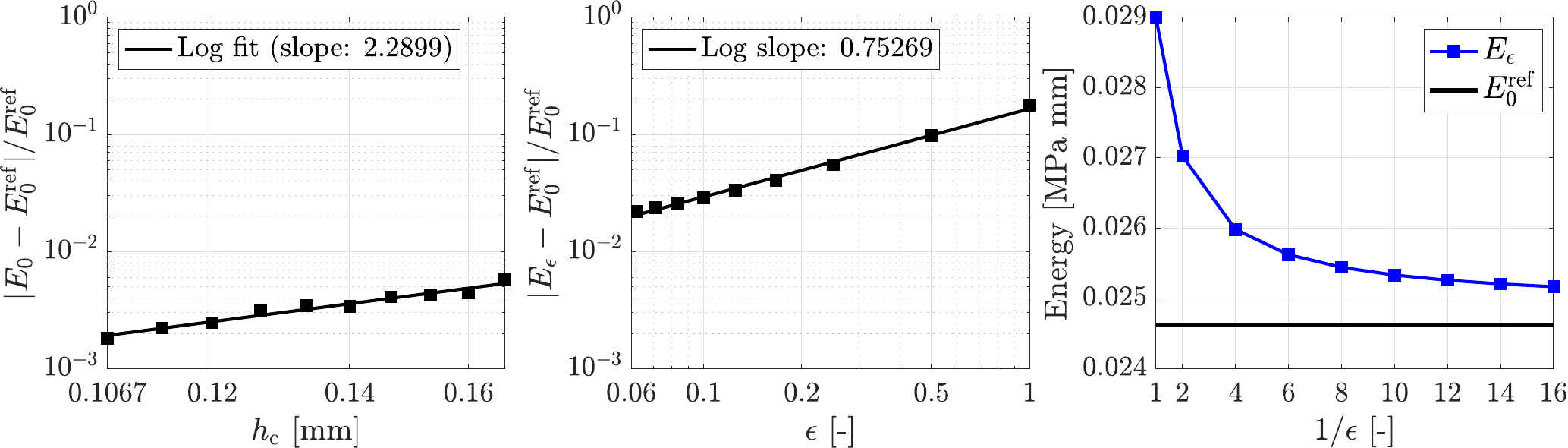}
    \put(-16cm,0cm){\small(a)}
    \put(-10.5cm,0cm){\small(b)}
    \put(-4.9cm,0cm){\small(c)}    
    \caption{Energy convergence and relative errors for the L-shaped specimen with an octet-truss microstructure.}   
    \label{fig:L_ene_relene_mesh}
\end{figure}

We proceed to assess the energy convergence numerically. Fig.~\ref{fig:L_ene_relene_mesh}a shows the convergence of the finite element simulations as the mesh size $h_\mathrm{c}$ decreases, with relative errors dropping from $0.57$\% at $h_\mathrm{c}=0.18$~mm to $0.18$\% at $h_\mathrm{c}=0.1$~mm. On the other hand, figures~\ref{fig:rotcube_ene_relene_mesh}b and~\ref{fig:rotcube_ene_relene_mesh}c show the convergence of the sequence of discrete energies~\eqref{R79mPV} to the continuum limit~\eqref{GQ4mav} as $\epsilon$ decreases, with relative errors ranging from $17.75$\% at $\epsilon=1$ to $2.2$\% at $1/\epsilon=16$. We emphasize that the larger errors compared to the two-dimensional examples are due to the coarser metamaterial; nevertheless, the convergence rate is well-defined. As in the previous example, the computational savings of the homogenized model are significant: $4\,763\,648$ beam elements and $4\,813\,120$ degrees of freedom for the largest discrete model ($1/\epsilon=16$), and $13\,489$ tetrahedral elements and $61\,353$ degrees of freedom for the finest continuum model, independent of $\epsilon$.

\section{Summary and concluding remarks}\label{sec:conc}

We have presented a computational assessment of homogenized models of mechanical metamaterials derived as the continuum limit of discrete energy functionals. Assuming a separation of scales between the lattice size and the macroscopic structure, it was shown numerically that the effective energies, corresponding to particular micropolar forms and descending naturally from arguments calculus of variations, yield solutions (displacements and rotations) that closely match the results of direct numerical simulations of discrete metastructures. Moreover, the sequence of discrete energies was shown numerically to converge to the continuum limit, with a well-defined convergence rate, depending on the structural geometry and boundary conditions. These numerical results were presented through various two-dimensional examples with a honeycomb microstructure and three-dimensional examples with an octet microstructure. In particular, the results verify the homogenization theory for mechanical metamaterials with axial and bending effects, elucidating the reliability--and convergence properties--of variational continuum models, suitable for simulating macro-scale metastructures.

The numerical results reveal that sharp corners and geometrical notches result in lower convergence rates. It is expected that in such cases, size effects come into play. The slower convergence is thus justified by the non-strict separation of scales and the fact that the zeroth-order model does not involve the lattice size. Indeed, we have shown in a separate work~\citep{Ulloa:2024} that second-order energy corrections, derived analytically through the framework of $\Gamma$-expansion~\citep{Anzellotti:1993}, are essential to capture size effects observed experimentally in octet-truss metastructures~\citep{Shaikeea2022}. It remains to formulate second-order models (or higher) suitable for numerical implementation in finite metastructures, likely as $\Gamma$-equivalent functionals~\citep{Braides:2008}.

\section*{Acknowledgements}

M.~Ortiz gratefully acknowledges the support of the Deutsche Forschungsgemeinschaft (DFG, German Research Foundation) {\sl via} project 211504053 - SFB 1060; project 441211072 - SPP 2256; and project 390685813 -  GZ 2047/1 - HCM. M.~P.~Ariza gratefully acknowledges financial support from Ministerio de Ciencia e Innovación under grant number PID2021-124869NB-I00. J.~Ulloa and J.~E.~Andrade acknowledge the support from the US ARO MURI program with Grant No. W911NF-19-1-0245. 

\begin{appendix}

\section{The discrete Fourier transform} \label{TwWqmj}

Let $(a_i)_{i=1}^n$ be a basis of $\mathbb{R}^n$ and $\mathcal{L} = \{ x(l) = \sum_{i=1}^n l^i a_i \, : \, l \in \mathbb{Z}^n \}$ the corresponding Bravais lattice. Let $f : \mathcal{L} \to \mathbb{R}$ be a real-valued lattice function. The discrete Fourier transform of $f$ is a complex function $\hat{f}(k)$ supported on the Brillouin zone $B$ in dual space given by
\begin{equation} \label{IYOTsT}
  \hat{f}(k) 
  = 
  V \sum_{l \in \mathbb{Z}^n}
  f(l) {\rm e}^{-i k \cdot x(l) } .
\end{equation}
where $V$ is the volume of the unit cell of the lattice. The inverse mapping is given by
\begin{equation} \label{aX0TEN}
  f(l) 
  = 
  \frac{1}{(2\pi)^n} \int_B
  \hat{f}(k) {\rm e}^{i k \cdot x(l) } dk .
\end{equation}
The convolution of two lattice functions $f(l)$, $g(l)$ is
\begin{equation} \label{YNU3CC}
  (f*g)(l) = V \sum_{l' \in \mathbb{Z}^n} f(l-l') g(l') ,
\end{equation}
whereupon the convolution theorem states that
\begin{equation} \label{m4IeYG}
  \widehat{f*g} = \hat{f} \hat{g} .
\end{equation}
In addition, the Parseval's identity states that
\begin{equation} \label{pB0rWJ}
  V \sum_{l \in \mathbb{Z}^n} f(l) g^*(l)
  = 
  \frac{1}{(2\pi)^n} \int_{B}
  \hat{f}(k) \hat{g}^*(k) dk
\end{equation}
which establishes an isometric isomorphism between $l^2$ and $L^2(B)$ (cf.~\cite{Ariza:2005} and references therein for further details).

\end{appendix}


\begin{thebibliography}{45}
\expandafter\ifx\csname natexlab\endcsname\relax\def\natexlab#1{#1}\fi
\providecommand{\url}[1]{\texttt{#1}}
\providecommand{\href}[2]{#2}
\providecommand{\path}[1]{#1}
\providecommand{\DOIprefix}{doi:}
\providecommand{\ArXivprefix}{arXiv:}
\providecommand{\URLprefix}{URL: }
\providecommand{\Pubmedprefix}{pmid:}
\providecommand{\doi}[1]{\href{http://dx.doi.org/#1}{\path{#1}}}
\providecommand{\Pubmed}[1]{\href{pmid:#1}{\path{#1}}}
\providecommand{\bibinfo}[2]{#2}
\ifx\xfnm\relax \def\xfnm[#1]{\unskip,\space#1}\fi
\bibitem[{Alavi et~al.(2021)Alavi, Ganghoffer, Reda and
  Sadighi}]{alavi2021construction}
\bibinfo{author}{Alavi, S.E.}, \bibinfo{author}{Ganghoffer, J.F.},
  \bibinfo{author}{Reda, H.}, \bibinfo{author}{Sadighi, M.},
  \bibinfo{year}{2021}.
\newblock \bibinfo{title}{Construction of micromorphic continua by
  homogenization based on variational principles}.
\newblock \bibinfo{journal}{Journal of the Mechanics and Physics of Solids}
  \bibinfo{volume}{153}, \bibinfo{pages}{104278}.
\bibitem[{Anzellotti and Baldo(1993)}]{Anzellotti:1993}
\bibinfo{author}{Anzellotti, G.}, \bibinfo{author}{Baldo, S.},
  \bibinfo{year}{1993}.
\newblock \bibinfo{title}{Asymptotic development by {$\Gamma$}-convergence}.
\newblock \bibinfo{journal}{Applied Mathematics and Optimization}
  \bibinfo{volume}{27}, \bibinfo{pages}{105--123}.
\bibitem[{Ariza et~al.(2024)Ariza, Conti and Ortiz}]{Ariza:2024}
\bibinfo{author}{Ariza, M.P.}, \bibinfo{author}{Conti, S.},
  \bibinfo{author}{Ortiz, M.}, \bibinfo{year}{2024}.
\newblock \bibinfo{title}{Homogenization and continuum limit of mechanical
  metamaterials}.
\newblock \bibinfo{journal}{Mechanics of Materials} \bibinfo{volume}{196},
  \bibinfo{pages}{105073}.
\bibitem[{Ariza and Ortiz(2005)}]{Ariza:2005}
\bibinfo{author}{Ariza, M.P.}, \bibinfo{author}{Ortiz, M.},
  \bibinfo{year}{2005}.
\newblock \bibinfo{title}{Discrete crystal elasticity and discrete dislocations
  in crystals}.
\newblock \bibinfo{journal}{Archive for Rational Mechanics and Analysis}
  \bibinfo{volume}{178}, \bibinfo{pages}{149--226}.
\bibitem[{Ashby(2006)}]{Ashby2006}
\bibinfo{author}{Ashby, M.F.}, \bibinfo{year}{2006}.
\newblock \bibinfo{title}{The properties of foams and lattices}.
\newblock \bibinfo{journal}{Philosophical Transactions of the Royal Society A:
  Mathematical, Physical and Engineering Sciences} \bibinfo{volume}{364},
  \bibinfo{pages}{15--30}.
\bibitem[{Ashby(2011)}]{ashby2011a}
\bibinfo{author}{Ashby, M.F.}, \bibinfo{year}{2011}.
\newblock \bibinfo{title}{Hybrid materials to expand the boundaries of
  material-property space}.
\newblock \bibinfo{journal}{Journal of the American Ceramic Society}
  \bibinfo{volume}{94}, \bibinfo{pages}{3-- 14}.
\bibitem[{Bauer et~al.(2014)Bauer, Hengsbach, Tesari, Schwaiger and
  Kraft}]{bauer2014a}
\bibinfo{author}{Bauer, J.}, \bibinfo{author}{Hengsbach, S.},
  \bibinfo{author}{Tesari, I.}, \bibinfo{author}{Schwaiger, R.},
  \bibinfo{author}{Kraft, O.}, \bibinfo{year}{2014}.
\newblock \bibinfo{title}{High-strength cellular ceramic composites with {3D}
  microarchitecture}.
\newblock \bibinfo{journal}{Proceedings of the National Academy of Sciences}
  \bibinfo{volume}{111}, \bibinfo{pages}{2453--2458}.
\bibitem[{Bauer et~al.(2016)Bauer, Schroer, Schwaiger and Kraft}]{bauer2016a}
\bibinfo{author}{Bauer, J.}, \bibinfo{author}{Schroer, A.},
  \bibinfo{author}{Schwaiger, R.}, \bibinfo{author}{Kraft, O.},
  \bibinfo{year}{2016}.
\newblock \bibinfo{title}{Approaching theoretical strength in glassy carbon
  nanolattices}.
\newblock \bibinfo{journal}{Nature Materials} \bibinfo{volume}{15},
  \bibinfo{pages}{438--443}.
\bibitem[{Bertoldi et~al.(2017)Bertoldi, Vitelli, Christensen and
  Van~Hecke}]{bertoldi2017}
\bibinfo{author}{Bertoldi, K.}, \bibinfo{author}{Vitelli, V.},
  \bibinfo{author}{Christensen, J.}, \bibinfo{author}{Van~Hecke, M.},
  \bibinfo{year}{2017}.
\newblock \bibinfo{title}{Flexible mechanical metamaterials}.
\newblock \bibinfo{journal}{Nature Reviews Materials} \bibinfo{volume}{2},
  \bibinfo{pages}{1--11}.
\bibitem[{Biswas et~al.(2020)Biswas, Poh and Shedbale}]{biswas2020micromorphic}
\bibinfo{author}{Biswas, R.}, \bibinfo{author}{Poh, L.H.},
  \bibinfo{author}{Shedbale, A.S.}, \bibinfo{year}{2020}.
\newblock \bibinfo{title}{A micromorphic computational homogenization framework
  for auxetic tetra-chiral structures}.
\newblock \bibinfo{journal}{Journal of the Mechanics and Physics of Solids}
  \bibinfo{volume}{135}, \bibinfo{pages}{103801}.
\bibitem[{Braides and Gelli(2004)}]{Braides:2004}
\bibinfo{author}{Braides, A.}, \bibinfo{author}{Gelli, M.S.},
  \bibinfo{year}{2004}.
\newblock \bibinfo{title}{The passage from discrete to continuous variational
  problems: A nonlinear homogenization process - continuum limits with bulk and
  surface energies}.
\newblock \bibinfo{journal}{Nonlinear Homogenization and Its Applications to
  Composites, Polycrystals and Smart Materials} \bibinfo{volume}{170},
  \bibinfo{pages}{45--63}.
\bibitem[{Braides and Truskinovsky(2008)}]{Braides:2008}
\bibinfo{author}{Braides, A.}, \bibinfo{author}{Truskinovsky, L.},
  \bibinfo{year}{2008}.
\newblock \bibinfo{title}{Asymptotic expansions by {$\Gamma$}-convergence}.
\newblock \bibinfo{journal}{Continuum Mechanics and Thermodynamics}
  \bibinfo{volume}{20}, \bibinfo{pages}{21--62}.
\bibitem[{Chen et~al.(1998)Chen, Huang and Ortiz}]{CHEN1998}
\bibinfo{author}{Chen, J.Y.}, \bibinfo{author}{Huang, Y.},
  \bibinfo{author}{Ortiz, M.}, \bibinfo{year}{1998}.
\newblock \bibinfo{title}{Fracture analysis of cellular materials: A strain
  gradient model}.
\newblock \bibinfo{journal}{Journal of the Mechanics and Physics of Solids}
  \bibinfo{volume}{46}, \bibinfo{pages}{789--828}.
\bibitem[{Cicalese et~al.(2009)Cicalese, DeSimone and Zeppieri}]{Cicalese:2009}
\bibinfo{author}{Cicalese, M.}, \bibinfo{author}{DeSimone, A.},
  \bibinfo{author}{Zeppieri, C.I.}, \bibinfo{year}{2009}.
\newblock \bibinfo{title}{Discrete-to-continuum limits for
  strain-alignment-coupled systems: Magnetostrictive solids, ferroelectric
  crystals and nematic elastomers}.
\newblock \bibinfo{journal}{Networks and Heterogeneous Media}
  \bibinfo{volume}{4}, \bibinfo{pages}{667--708}.
\bibitem[{Danesh et~al.(2023)Danesh, Brepols and Reese}]{danesh2023}
\bibinfo{author}{Danesh, H.}, \bibinfo{author}{Brepols, T.},
  \bibinfo{author}{Reese, S.}, \bibinfo{year}{2023}.
\newblock \bibinfo{title}{Challenges in two-scale computational homogenization
  of mechanical metamaterials}.
\newblock \bibinfo{journal}{PAMM} \bibinfo{volume}{23},
  \bibinfo{pages}{e202200139}.
\bibitem[{Deshpande et~al.(2001a)Deshpande, Ashby and Fleck}]{deshpande2001a}
\bibinfo{author}{Deshpande, V.S.}, \bibinfo{author}{Ashby, M.F.},
  \bibinfo{author}{Fleck, N.A.}, \bibinfo{year}{2001}a.
\newblock \bibinfo{title}{Foam topology: bending versus stretching dominated
  architectures}.
\newblock \bibinfo{journal}{Acta Materialia} \bibinfo{volume}{49},
  \bibinfo{pages}{1035--1040}.
\bibitem[{Deshpande et~al.(2001b)Deshpande, Fleck and Ashby}]{deshpande2001b}
\bibinfo{author}{Deshpande, V.S.}, \bibinfo{author}{Fleck, N.A.},
  \bibinfo{author}{Ashby, M.F.}, \bibinfo{year}{2001}b.
\newblock \bibinfo{title}{Effective properties of the octet-truss lattice
  material}.
\newblock \bibinfo{journal}{Journal of the Mechanics and Physics of Solids}
  \bibinfo{volume}{49}, \bibinfo{pages}{1747--1769}.
\bibitem[{Dos~Reis and Ganghoffer(2012)}]{dos2012construction}
\bibinfo{author}{Dos~Reis, F.}, \bibinfo{author}{Ganghoffer, J.F.},
  \bibinfo{year}{2012}.
\newblock \bibinfo{title}{Construction of micropolar continua from the
  asymptotic homogenization of beam lattices}.
\newblock \bibinfo{journal}{Computers \& Structures} \bibinfo{volume}{112},
  \bibinfo{pages}{354--363}.
\bibitem[{Eringen(1966)}]{Eringen:1966}
\bibinfo{author}{Eringen, A.C.}, \bibinfo{year}{1966}.
\newblock \bibinfo{title}{Linear theory of micropolar elasticity}.
\newblock \bibinfo{journal}{Journal of Mathematics and Mechanics}
  \bibinfo{volume}{15}, \bibinfo{pages}{909--923}.
\bibitem[{Eringen and Suhubi(1964)}]{Eringen:1964}
\bibinfo{author}{Eringen, A.C.}, \bibinfo{author}{Suhubi, E.S.},
  \bibinfo{year}{1964}.
\newblock \bibinfo{title}{{Nonlinear theory of simple micro-elastic solids-I}}.
\newblock \bibinfo{journal}{International Journal of Engineering Science}
  \bibinfo{volume}{2}, \bibinfo{pages}{189--203}.
\bibitem[{Espa{\~n}ol et~al.(2013)Espa{\~n}ol, Kochmann, Conti and
  Ortiz}]{Espanol:2013}
\bibinfo{author}{Espa{\~n}ol, M.I.}, \bibinfo{author}{Kochmann, D.M.},
  \bibinfo{author}{Conti, S.}, \bibinfo{author}{Ortiz, M.},
  \bibinfo{year}{2013}.
\newblock \bibinfo{title}{A {$\Gamma$}-convergence analysis of the
  quasicontinuum method.}
\newblock \bibinfo{journal}{SIAM Multiscale Modeling {\&} Simulation}
  \bibinfo{volume}{11}, \bibinfo{pages}{766--794}.
\bibitem[{Fleck et~al.(2010)Fleck, Deshpande and Ashby}]{Fleck2010}
\bibinfo{author}{Fleck, N.A.}, \bibinfo{author}{Deshpande, V.S.},
  \bibinfo{author}{Ashby, M.F.}, \bibinfo{year}{2010}.
\newblock \bibinfo{title}{Micro-architectured materials: Past, present and
  future}.
\newblock \bibinfo{journal}{Proceedings of the Royal Society A: Mathematical,
  Physical and Engineering Sciences} \bibinfo{volume}{466},
  \bibinfo{pages}{2495--2516}.
\bibitem[{Fran{\c{c}}ois et~al.(2021)Fran{\c{c}}ois, Schevenels, Dooms, Jansen,
  Wambacq, Lombaert, Degrande and De~Roeck}]{franccois2021stabil}
\bibinfo{author}{Fran{\c{c}}ois, S.}, \bibinfo{author}{Schevenels, M.},
  \bibinfo{author}{Dooms, D.}, \bibinfo{author}{Jansen, M.},
  \bibinfo{author}{Wambacq, J.}, \bibinfo{author}{Lombaert, G.},
  \bibinfo{author}{Degrande, G.}, \bibinfo{author}{De~Roeck, G.},
  \bibinfo{year}{2021}.
\newblock \bibinfo{title}{Stabil: An educational matlab toolbox for static and
  dynamic structural analysis}.
\newblock \bibinfo{journal}{Computer Applications in Engineering Education}
  \bibinfo{volume}{29}, \bibinfo{pages}{1372--1389}.
\bibitem[{Greer and Deshpande(2019)}]{Greer:2019}
\bibinfo{author}{Greer, J.R.}, \bibinfo{author}{Deshpande, V.S.},
  \bibinfo{year}{2019}.
\newblock \bibinfo{title}{Three-dimensional architected materials and
  structures: Design, fabrication, and mechanical behavior}.
\newblock \bibinfo{journal}{MRS Bulletin} \bibinfo{volume}{44},
  \bibinfo{pages}{750–757}.
\bibitem[{Gu and Greer(2015)}]{gu2015a}
\bibinfo{author}{Gu, X.}, \bibinfo{author}{Greer, J.R.}, \bibinfo{year}{2015}.
\newblock \bibinfo{title}{Ultra-strong architected {Cu} meso-lattices}.
\newblock \bibinfo{journal}{Extreme Mechanics Letters} \bibinfo{volume}{2},
  \bibinfo{pages}{7--14}.
\bibitem[{Jin and Espinosa(2024)}]{Jin2024}
\bibinfo{author}{Jin, H.}, \bibinfo{author}{Espinosa, H.D.},
  \bibinfo{year}{2024}.
\newblock \bibinfo{title}{Mechanical metamaterials fabricated from
  self-assembly: A perspective}.
\newblock \bibinfo{journal}{Journal of Applied Mechanics, Transactions ASME}
  \bibinfo{volume}{91}.
\bibitem[{Kochmann et~al.(2019)Kochmann, Hopkins and Valdevit}]{kochmann:2019}
\bibinfo{author}{Kochmann, D.M.}, \bibinfo{author}{Hopkins, J.B.},
  \bibinfo{author}{Valdevit, L.}, \bibinfo{year}{2019}.
\newblock \bibinfo{title}{Multiscale modeling and optimization of the mechanics
  of hierarchical metamaterials}.
\newblock \bibinfo{journal}{MRS Bulletin} \bibinfo{volume}{44},
  \bibinfo{pages}{773--781}.
\bibitem[{Kuszczak et~al.(2023)Kuszczak, Azam, Bessa, Tan and
  Bosi}]{kuszczak2023}
\bibinfo{author}{Kuszczak, I.}, \bibinfo{author}{Azam, F.I.},
  \bibinfo{author}{Bessa, M.A.}, \bibinfo{author}{Tan, P.J.},
  \bibinfo{author}{Bosi, F.}, \bibinfo{year}{2023}.
\newblock \bibinfo{title}{Bayesian optimisation of hexagonal honeycomb
  metamaterial}.
\newblock \bibinfo{journal}{Extreme Mechanics Letters} \bibinfo{volume}{64},
  \bibinfo{pages}{102078}.
\bibitem[{Lu et~al.(2022)Lu, Hsieh, Huang, Zhang, Lin, Shen, Chen and
  Zhang}]{LU2022}
\bibinfo{author}{Lu, C.}, \bibinfo{author}{Hsieh, M.}, \bibinfo{author}{Huang,
  Z.}, \bibinfo{author}{Zhang, C.}, \bibinfo{author}{Lin, Y.},
  \bibinfo{author}{Shen, Q.}, \bibinfo{author}{Chen, F.},
  \bibinfo{author}{Zhang, L.}, \bibinfo{year}{2022}.
\newblock \bibinfo{title}{Architectural design and additive manufacturing of
  mechanical metamaterials: A review}.
\newblock \bibinfo{journal}{Engineering} \bibinfo{volume}{17},
  \bibinfo{pages}{44--63}.
\bibitem[{dal Maso(1993)}]{dalmaso:1993}
\bibinfo{author}{dal Maso, G.}, \bibinfo{year}{1993}.
\newblock \bibinfo{title}{An introduction to {$\Gamma$}-convergence}.
\newblock Progress in Nonlinear Differential Equations and their Applications,
  8, \bibinfo{publisher}{Birkh\"auser Boston Inc.}, \bibinfo{address}{Boston,
  MA}.
\bibitem[{Meza et~al.(2015)Meza, Zelhofer, Clarke, Mateos, Kochmann and
  Greer}]{meza2015a}
\bibinfo{author}{Meza, L.R.}, \bibinfo{author}{Zelhofer, A.J.},
  \bibinfo{author}{Clarke, N.}, \bibinfo{author}{Mateos, A.J.},
  \bibinfo{author}{Kochmann, D.M.}, \bibinfo{author}{Greer, J.R.},
  \bibinfo{year}{2015}.
\newblock \bibinfo{title}{Resilient {3D} hierarchical architected
  metamaterials}.
\newblock \bibinfo{journal}{Proceedings of the National Academy of Sciences}
  \bibinfo{volume}{112}, \bibinfo{pages}{11502--11507}.
\bibitem[{Misra et~al.(2020)Misra, Nejadsadeghi, De~Angelo and
  Placidi}]{misra2020chiral}
\bibinfo{author}{Misra, A.}, \bibinfo{author}{Nejadsadeghi, N.},
  \bibinfo{author}{De~Angelo, M.}, \bibinfo{author}{Placidi, L.},
  \bibinfo{year}{2020}.
\newblock \bibinfo{title}{Chiral metamaterial predicted by granular
  micromechanics: verified with {1D} example synthesized using additive
  manufacturing}.
\newblock \bibinfo{journal}{Continuum Mechanics and Thermodynamics}
  \bibinfo{volume}{32}, \bibinfo{pages}{1497--1513}.
\bibitem[{Montemayor et~al.(2015)Montemayor, Chernow and
  Greer}]{Montemayor2015}
\bibinfo{author}{Montemayor, L.}, \bibinfo{author}{Chernow, V.},
  \bibinfo{author}{Greer, J.R.}, \bibinfo{year}{2015}.
\newblock \bibinfo{title}{Materials by design: Using architecture in material
  design to reach new property spaces}.
\newblock \bibinfo{journal}{MRS Bulletin} \bibinfo{volume}{40},
  \bibinfo{pages}{1122--1129}.
\bibitem[{Neff et~al.(2020)Neff, Eidel, d’Agostino and
  Madeo}]{neff2020identification}
\bibinfo{author}{Neff, P.}, \bibinfo{author}{Eidel, B.},
  \bibinfo{author}{d’Agostino, M.V.}, \bibinfo{author}{Madeo, A.},
  \bibinfo{year}{2020}.
\newblock \bibinfo{title}{Identification of scale-independent material
  parameters in the relaxed micromorphic model through model-adapted first
  order homogenization}.
\newblock \bibinfo{journal}{Journal of Elasticity} \bibinfo{volume}{139},
  \bibinfo{pages}{269--298}.
\bibitem[{Phlipot and Kochmann(2019)}]{PHLIPOT2019}
\bibinfo{author}{Phlipot, G.P.}, \bibinfo{author}{Kochmann, D.M.},
  \bibinfo{year}{2019}.
\newblock \bibinfo{title}{A quasicontinuum theory for the nonlinear mechanical
  response of general periodic truss lattices}.
\newblock \bibinfo{journal}{Journal of the Mechanics and Physics of Solids}
  \bibinfo{volume}{124}, \bibinfo{pages}{758--780}.
\bibitem[{Roko{\v{s}} et~al.(2019)Roko{\v{s}}, Ameen, Peerlings and
  Geers}]{rokovs2019micromorphic}
\bibinfo{author}{Roko{\v{s}}, O.}, \bibinfo{author}{Ameen, M.M.},
  \bibinfo{author}{Peerlings, R.H.J.}, \bibinfo{author}{Geers, M.G.D.},
  \bibinfo{year}{2019}.
\newblock \bibinfo{title}{Micromorphic computational homogenization for
  mechanical metamaterials with patterning fluctuation fields}.
\newblock \bibinfo{journal}{Journal of the Mechanics and Physics of Solids}
  \bibinfo{volume}{123}, \bibinfo{pages}{119--137}.
\bibitem[{Rosário et~al.(2015)Rosário, Lilleodden, Waleczek, Kubrin, Petrov,
  Dyachenko, Sabisch, Nielsch, Huber, Eich and Schneider}]{ros2015a}
\bibinfo{author}{Rosário, J.J.}, \bibinfo{author}{Lilleodden, E.T.},
  \bibinfo{author}{Waleczek, M.}, \bibinfo{author}{Kubrin, R.},
  \bibinfo{author}{Petrov, A.}, \bibinfo{author}{Dyachenko, P.N.},
  \bibinfo{author}{Sabisch, J.E.C.}, \bibinfo{author}{Nielsch, K.},
  \bibinfo{author}{Huber, N.}, \bibinfo{author}{Eich, M.},
  \bibinfo{author}{Schneider, G.A.}, \bibinfo{year}{2015}.
\newblock \bibinfo{title}{Self-assembled ultra high strength, ultra stiff
  mechanical metamaterials based on inverse opals}.
\newblock \bibinfo{journal}{Advanced Engineering Materials}
  \bibinfo{volume}{17}, \bibinfo{pages}{1420--1424}.
\bibitem[{Rys et~al.(2014)Rys, Valdevit, Schaedler, Jacobsen, Carter and
  Greer}]{rys2014a}
\bibinfo{author}{Rys, J.}, \bibinfo{author}{Valdevit, L.},
  \bibinfo{author}{Schaedler, T.A.}, \bibinfo{author}{Jacobsen, A.J.},
  \bibinfo{author}{Carter, W.B.}, \bibinfo{author}{Greer, J.R.},
  \bibinfo{year}{2014}.
\newblock \bibinfo{title}{Fabrication and deformation of metallic glass
  micro-lattices}.
\newblock \bibinfo{journal}{Advanced Engineering Materials}
  \bibinfo{volume}{16}, \bibinfo{pages}{889--896}.
\bibitem[{Shaikeea et~al.(2022)Shaikeea, Cui, O’Masta, Zheng and
  Deshpande}]{Shaikeea2022}
\bibinfo{author}{Shaikeea, A.J.D.}, \bibinfo{author}{Cui, H.},
  \bibinfo{author}{O’Masta, M.}, \bibinfo{author}{Zheng, X.R.},
  \bibinfo{author}{Deshpande, V.S.}, \bibinfo{year}{2022}.
\newblock \bibinfo{title}{The toughness of mechanical metamaterials}.
\newblock \bibinfo{journal}{Nature Materials} \bibinfo{volume}{21},
  \bibinfo{pages}{297--304}.
\bibitem[{Suh and Sun(2019)}]{suh2019open}
\bibinfo{author}{Suh, H.S.}, \bibinfo{author}{Sun, W.C.}, \bibinfo{year}{2019}.
\newblock \bibinfo{title}{An open-source {FEniCS} implementation of a phase
  field fracture model for micropolar continua}.
\newblock \bibinfo{journal}{International Journal for Multiscale Computational
  Engineering} \bibinfo{volume}{17}.
\bibitem[{Timoshenko and Young(1965)}]{Timoshenko:1965}
\bibinfo{author}{Timoshenko, S.P.}, \bibinfo{author}{Young, D.H.},
  \bibinfo{year}{1965}.
\newblock \bibinfo{title}{Theory of Structures}.
\newblock \bibinfo{edition}{2nd} ed., \bibinfo{publisher}{McGraw-Hill Book
  Company}, \bibinfo{address}{New York}.
\bibitem[{Ulloa et~al.(2024)Ulloa, Ariza, Andrade and Ortiz}]{Ulloa:2024}
\bibinfo{author}{Ulloa, J.}, \bibinfo{author}{Ariza, M.P.},
  \bibinfo{author}{Andrade, J.E.}, \bibinfo{author}{Ortiz, M.},
  \bibinfo{year}{2024}.
\newblock \bibinfo{title}{Fracture and size effect in mechanical
  metamaterials}.
\newblock \bibinfo{journal}{Journal of the Mechanics and Physics of Solids}
  \bibinfo{volume}{193}, \bibinfo{pages}{105860}.
\bibitem[{Vigliotti and Pasini(2012)}]{Vigliotti2012}
\bibinfo{author}{Vigliotti, A.}, \bibinfo{author}{Pasini, D.},
  \bibinfo{year}{2012}.
\newblock \bibinfo{title}{Stiffness and strength of tridimensional periodic
  lattices}.
\newblock \bibinfo{journal}{Computer Methods in Applied Mechanics and
  Engineering} \bibinfo{volume}{229-232}, \bibinfo{pages}{27--43}.
\bibitem[{Weinberg(2023)}]{Weinberg:2023}
\bibinfo{author}{Weinberg, K.}, \bibinfo{year}{2023}.
\newblock \bibinfo{title}{Data-driven finite element computation of
  microstructured materials}.
\newblock \bibinfo{journal}{PAMM} \bibinfo{volume}{23},
  \bibinfo{pages}{e202300285}.
\bibitem[{Zhang and Bhattacharya(2024)}]{zhang2024}
\bibinfo{author}{Zhang, Y.}, \bibinfo{author}{Bhattacharya, K.},
  \bibinfo{year}{2024}.
\newblock \bibinfo{title}{Iterated learning and multiscale modeling of
  history-dependent architectured metamaterials}.
\newblock \bibinfo{journal}{Mechanics of Materials} \bibinfo{volume}{197},
  \bibinfo{pages}{105090}.

\end{thebibliography}
\end{document}